\documentclass[useAMS,usenatbib,psfig,times]{mn2e}
\usepackage{psfig}
\usepackage{tabularx}
\usepackage{graphicx}

\title[CENSORS: Spectroscopic observations]{A Combined EIS-NVSS Survey Of Radio Sources (CENSORS) III: Spectroscopic observations}

\author[M.H.~Brookes
  \etal]{M.H.~Brookes$^{1,2}$,P.N.~Best$^1$\thanks{Email: pnb@roe.ac.uk},
  J.A.~Peacock$^1$, H.J.A.~R\"{o}ttgering$^3$, J.S.~Dunlop$^1$\\  
$^1$ SUPA\thanks{Scottish Universities Physics Alliance}, Institute for Astronomy, University of Edinburgh, Blackford Hill,
Edinburgh EH9 3HJ\\
$^2$ Jet Propulsion Laboratory, 4800 Oak Grove Dr., Pasadena, CA 91101, USA\\
$^3$ Sterrewacht Leiden, Postbus 9513, 2300 RA Leiden, the Netherlands\\
}
\pagerange{\pageref{firstpage}--\pageref{lastpage}}
\pubyear{2004}
\begin{document}
\label{firstpage}

\def\lesssim{\mathrel{\hbox{\rlap{\hbox{\lower4pt\hbox{$\sim$}}}\hbox{$<$}}}}
\def\gtrsim{\mathrel{\hbox{\rlap{\hbox{\lower4pt\hbox{$\sim$}}}\hbox{$>$}}}}
\def\subsun{\mbox{$_{\normalsize\odot}$}}
\def\deg{\hbox{$^\circ$}}
\def\arcs{\ifmmode {^{\scriptscriptstyle\prime\prime}}
          \else $^{\scriptscriptstyle\prime\prime}$\fi}
\def\arcm{\ifmmode {^{\scriptscriptstyle\prime}}
          \else $^{\scriptscriptstyle\prime}$\fi}
\def\squig{$\sim\!\!$}
\def\Hb{\ifmmode {\rm H\beta}
          \else \rm H$\beta$\fi}
\def\Hg{\ifmmode {H\gamma}
          \else H$\gamma$\fi}
\def\Hd{\ifmmode {H\delta}
          \else H$\delta$\fi}
\def\vvmax{$V/V_{\rm max}$}
\def\etal{et~al.}

\def\nii{[N\,{\footnotesize II}]}
\def\oii{[O\,{\footnotesize II}]}
\def\oiii{[\rm O\,{\footnotesize \rm III}]}
\def\Ha{H$\alpha$}
\def\lya{Ly$\alpha$}
\def\14{\rm 1.4\,GHz}
\def\27{\rm 2.7\,GHz}
\def\whz1{$\,\rm W\,Hz^{-1}$}
\def\kms1{$\,\rm km\,s^{-1}$}

\maketitle

\begin{abstract}
\noindent 
The Combined EIS-NVSS Survey Of Radio Sources (CENSORS) is a \14 radio
survey selected from the NRAO VLA Sky Survey (NVSS) and complete to a
flux-density of 7.2\,mJy. It targets the ESO Imaging Survey (EIS) Patch D,
which is a 3 by 2 square degree field centred on 09 51 36.0 $-21$ 00 00
(J2000).  This paper presents the results of spectroscopic observations of
143 of the 150 CENSORS sources.  The primary motivation for these
observations is to achieve sufficient spectroscopic completeness so that
the sample may be used to investigate the evolution of radio sources.
 
The observations result in secure spectroscopic redshifts for 63\% of the
sample and likely redshifts (based on a single emission line, for example)
for a further 8\%.  Following the identification of the quasars and
star-forming galaxies in the CENSORS sample, estimated redshifts are
calculated for the remainder of the sample via the $K$--$z$ relation for
radio galaxies.

Comparison of the redshift distribution of the CENSORS radio sources to
distributions predicted by the various radio luminosity function evolution
models of \citeauthor{DP90}, results in no good match. This demonstrates
that this sample can be used to expand upon previous work in that field.
\end{abstract}


\section{Introduction}

The high redshift evolution of the radio luminosity function (RLF) has
been speculated upon for several decades.  In 1990, \citeauthor{DP90}
presented an investigation of both steep and flat spectrum, bright radio
sources at \27.  This work firmly established an increase, by 3 orders of
magnitude, in the comoving number density of sources between redshift
$\simeq$~0 and 2.  At higher redshifts they presented evidence for a
decline (or high redshift ``cut-off'') in the number density of sources
with both steep and flat spectrum.  However this result was dependent on
the accuracy of photometric redshifts ascribed to sources from the
faintest of their radio selected samples, the Parkes Selected Regions
(\citeauthor{Downes86} 1986 and \citeauthor{Dunlop89} 1989).  They were
not able to measure the extent of the decline with redshift, only to show
that the best fitting models predicted a decline in most cases. Thus,
given the redshift uncertainty, it is possible that the number density of
these sources doesn't actually decline until higher redshifts.  The
difference between a number density turnover at redshifts of 2 and 4
represents roughly 1.9\,Gyrs in the history of the Universe.

Establishing the presence (or absence) of a high redshift cut-off is of
interest because it provides information about the timescales over which
the population of radio sources is built-up in the early Universe. This is
important because the history of AGN (active galactic nuclei) activity is
closely related to that of the `normal' galaxy population. All nearby
galaxies contain a supermassive black hole with mass roughly proportional
to the galaxy bulge mass \citep{KG01}. Since radio-loud AGN typically
reside in the most massive old elliptical galaxies \citep{BLR98,Best05},
which host the most massive central black holes, the build-up of the radio
source population offers a unique probe of the early evolution of the
upper end of the black hole mass function. This is especially interesting
given the possibility that the AGN themselves form part of a feedback
system in which their activity affects the growth of the black hole and
host galaxy (see, for example, \citeauthor{spring05} 2005 and
\citeauthor{Best06} 2006).

The evolution of the radio-loud AGN population can also be related to that
of AGN selected in other wavebands. There is a sharp cut-off in the
density of optically-selected quasi-stellar objects (QSOs) at redshifts
greater than 2.1 (\citeauthor{2QZ} 2000, \citeauthor{Fan01} et al. 2001,
\citeauthor{Wolf03} et al. 2003, \citeauthor{Fan04} et al. 2004 and
references therein). The flat spectrum radio sources broadly correspond to
radio--loud QSOs (as implied by \citeauthor{barthel89} 1989 and
\citeauthor{antonucci93} 1993). Therefore the evolution of flat and steep
spectrum radio sources may be used to test/probe the relation between
radio-quiet QSOs and radio-loud QSOs (quasars), and between quasars and
radio galaxies.  In addition, the advent of high resolution X-ray
telescopes has enabled the density of X-ray active galaxies to be studied.
These sources show a variation in the redshift at which the population
peaks in density as a function of luminosity \citep{Cow03}. At the heart
of all of these studies is a desire to understand what the various types
of AGN are that give rise to the sources seen at different wavelengths and
how they evolve to produce the observed population behaviour.

Some progress has been made since 1990, largely in the study of flat
spectrum radio sources.  In 2000, Jarvis and Rawlings\nocite{JR00}, found
some evidence for a shallow decline, with redshift, in the number density
of flat-spectrum radio sources at high redshifts (refuting an earlier
claim of a sharp cut-off, in a similar sample, by \citeauthor{Shav96}
1996).  Subsequently, \citet{wall05} provided a more rigorous analysis of
the behaviour of radio-loud quasars and showed that their densities are
consistent with the behaviour of optical QSOs, i.e., a decline in number
density at $z > 3$.

In 2001 Jarvis et al.\nocite{jarvis_st01} investigated the behaviour of
steep spectrum sources, finding both a shallow decline or levelling off of
the number density, between $z \simeq$~2.5 and 4.5, to be consistent with
the data.  This sample lacked sufficient depth to precisely probe the
evolution of the high redshift sources.  Waddington et
al. (2001)\nocite{Waddington01} used deep radio observations to
investigate luminosity function evolution and were able to discount some
of the models from Dunlop and Peacock, but this sample did not have
sufficient volume to improve measurements of the number density of the
most powerful sources.  Interestingly, they did find evidence that the
space density of mJy radio sources began to turn-over at lower redshifts
than that of samples with a much brighter flux-density limit. Thus, whilst
it is clear that the number density of luminous steep spectrum sources
increases up to redshifts beyond 2, none of these surveys, due to lack of
depth or volume, has provided sufficient information about the powerful
sources at redshifts beyond this.

Over the last five years the Combined EIS-NVSS Survey Of Radio Sources
(CENSORS; \citeauthor{CENSORS1} 2003; hereafter Paper 1) has been
developed with one of the primary scientific goals being to investigate
the RLF at redshifts $\gtrsim$~2.  The sample contains 150 sources of
radio flux density (at 1.4\,GHz) greater than 3.8\,mJy.  Of these 137 are
deemed to be complete to a flux-density limit of 7.2\,mJy.  As argued in
Paper 1, CENSORS is of optimal depth to maximise information for the radio
sources close to the break in the radio luminosity function at redshifts
of $\simeq$~2.5, and is thus ideal for investigating radio source number
densities at these redshifts.  In order to carry out such investigations,
accurate redshifts are essential for a large fraction, preferably all, of
the sources in the sample.  \citeauthor{CENSORS2} (2006; hereafter Paper
2) described the follow-up imaging of the CENSORS sample, providing 92\%
of the sample with a clear host galaxy candidate.  Only a few objects
remain without any detections at all, some of which involve complications
which may be solved by further follow-up in the radio.  In order to
achieve the high degree of spectroscopic completeness required, these host
galaxy candidates have subsequently been targetted in a program of
spectroscopic observations and it is these data which are described in
this paper.

The layout of this paper is as follows: in Section \ref{observations} the
observing strategy is described, followed by the data reduction techniques
in Section \ref{datareduction}.  Section \ref{results} presents the
results of the spectroscopy, and Section \ref{kzsection} describes the use
of the $K$--$z$ relation for radio galaxies to estimate the redshifts for
those sources which lack a spectroscopic redshift.  Section \ref{NZ}
presents the redshift distribution for CENSORS and compares it to the
predictions of \citet{DP90}.  Throughout this paper the following
cosmological parameters are adopted: $H_{0} = 70\,{\rm
kms}^{-1}\rm{Mpc}^{-1}$, $\Omega_{M} = 0.3$ and $\Omega_{\Lambda} = 0.7$.
Please note, if comparing to the work of \citet{DP90}, that they assumed
an Einstein--de Sitter cosmology.

\begin{scriptsize}
\begin{table*}

\begin{tabular}{|l|l|l|l|l|l|l|l|} \hline 
Run       	&Date          	&Telescope (Instr.)    		&Grism/Grating	&Central $\lambda$	&Resolution	&Typical	&Strategy 		\\ 
		&		&				&		&(\AA)			&(\AA)		&Seeing	($\arcs$)&			\\ \hline
1a		&20000301       &AAT		(2dF)		&270R		&6503			&10		&1.5		&4 pointings 		\\
		&		&				&		&			&		&		&of 3*1800s 		\\
1b		&20010307	&AAT		(2dF)		&270R		&6503			&10		&1.5		&2 pointings 		\\
		&		&				&		&			&		&		&of 3*1800s 		\\
2		&20020212 - 15 	&VLT		(FORS1)		&300V		&5850			&11.1		&0.7-1.5	&Up to 3 x 20 		\\
		&		&				&		&			&		&		&min exposures.		\\
3		&20030204 - 08 	&ESO 3.6m		(EFOSC2)	&Gr\#6		&5965			&12.9		&1.0		&Up to 3 x 30 		\\
		&		&				&		&			&		&		&min exposures.		\\
4		&20030226 - 28	&VLT		(FORS1)		&300V		&5850			&11.1		&0.7-1.5	&Up to 3 x 30 		\\
		&		&				&		&			&		&		&min exposures.		\\
5		&20050206	&VLT		(FORS1)		&300V		&5850			&11.1		&0.7-1.5	&600-1800s 		\\
		&		&				&		&			&		&		&exposures		\\
6		&20050413	&WHT		(ISIS)		&R316R/R158B	&6500/3600		&12/13		&1.4		&900s			\\
7		&20060223	&VLT		(FORS1)		&300V		&5850			&11.1		&0.7-1.5	&Up to 3 x 30		\\
		&		&				&		&			&		&		&min exposures		\\
8		&20060226-28	&VLT		(FORS2)		&300I		&8600			&11.3		&1.0-1.5	&1 to 4 x 20  		\\
		&		&				&		&			&		&		&min exposures  	\\
\hline
\end{tabular}
\caption{Details of the spectroscopic observations for
CENSORS.\label{spec_obs_details} }

\end{table*}

\end{scriptsize}

\section{Observations}
\label{observations}
The CENSORS sources detected in the $I$--band have magnitudes in the range
$I \simeq$~15 to $I \simeq$~23 (the limit of the ESO Imaging Survey Wide
survey). $K$--band follow up has revealed sources as faint as $K$ = 20.5.
Given the range in optical brightness, the following strategy (using five
telescopes in eight observing runs) was adopted for spectroscopy (details
of the observing runs are given in Table \ref{spec_obs_details}).

Multi-Object fibre Spectroscopy (MOS) using the Two Degree Field
Spectrograph (2dF) on the Anglo-Australian Telescope (AAT) was employed
first, as this allows the spectra of many targets to be obtained
simultaneously.  The 2dF consists of two spectrographs, each of which
accesses 200 optical fibres which are placed on target sources, or blank
sky for subtraction purposes, over a two degree field of view. The vast
majority of the CENSORS field can therefore be covered in 4 pointings.
Due to the small number of CENSORS sources per pointing, only one of the
spectrographs was used.  For each pointing, the 5400s observations were
split into three 1800s exposures. In addition, a flat field and an arc
were taken through the same fibre configuration, plus three 300s offset
`blank' sky frames (at three different positions) which were used to
calibrate the optical throughput of the fibres.  The first attempt was
made in March 2000 but poor weather affected two of the pointings so these
observations were repeated the following year (unfortunately, these were
also weather affected).

Due to the difficulty of obtaining good sky subtraction, a 4m class
telescope with fibre observations is sensitive only to the optically
brightest sources, or those with bright emission lines.  So, while the
majority of the CENSORS sources could be targetted with 2dF, redshifts
were only found for $\simeq$\,30\%.

Long slit spectroscopy (LSS) was then used to target those sources without
a measured redshift from the 2dF observations.  The ESO 3.6m telescope at
La Silla was used in 2003 to target intermediately-bright sources (in the
optical) with a 1.5\arcs~~slit.  Up to three exposures of 30 minutes each
were taken, with the later exposures being stopped if a redshift was
immediately apparent in the first data.

The optically-faintest sources (generally those which were only detected
in the $K$--band) were targetted with FORS1 on the 8m class ESO Very Large
Telescope (VLT), in 2002, 2003 and 2006, using a 1\arcs\--1.5\arcs\,slit.
Further service mode observations were made in 2005 using the VLT and the
William Herschel Telescope (WHT).  A subset of sources for which no
redshift was obtained were targetted at redder wavelengths with FORS2 in
2006. Full details of the observations are provided in
Tables~\ref{spec_obs_details} and A1.

\section{Data Reduction}
\label{datareduction}

\subsection{Multi-object spectroscopy}
\label{2dfdr}

The process of 2dF data reduction involves: removal of bad pixels and bias
subtraction; extraction of the spectra from the raw frame; wavelength
calibration with reference to the extracted arc observation; fibre
throughput estimation and correction via offset sky images; sky
subtraction, using the median sky as calculated from sky fibres; and,
finally, combination of multiple exposures.  No flux calibration was
available for these data.

The 2dF data reduction pipeline (2dFdr; see {\tt
http://www.aao.gov.au/2df/software.html}) was used for the initial steps
of this process, up to and including the fibre extraction. It was not used
thereafter, because the pipeline results produced poor quality spectra, in
which only the brightest sources could be identified
spectroscopically. Many of the spectra were over or under subtracted in
relation to the sky spectrum. This over-/under-subtraction varied with
position on the extracted image indicating that it was likely due to
variation of the sky background over the field of view. It should be noted
that this is a particular problem for galaxies such as those in CENSORS
because they are so faint, making sky subtraction errors particularly
problematic.

To avoid this problem, an improved sky subtraction code was developed. The
three pipeline-produced fibre-extracted images (which each contain about
200 rows corresponding to the fibres observed) for both the target and the
offset sky were combined in \textsc{IRAF} with cosmic ray rejection. The
combined offset sky image was then used to determine the relative fibre
throughput, and the combined target frame was corrected for this. Sky
subtraction was then carried out by assuming that the sky varied, with
fibre number, as a low order polynomial (of order 2) plus a distortion,
introduced by the throughput and scattered light term.  The sky fibre
counts were fitted as a function of fibre number for all wavelengths and
subtracted from all fibres.  The median of the residual background (taken
from the sky subtracted sky fibres) was then subtracted, for all pixels in
wavelength, to account for remaining distortion. This dramatically
improved the background subtraction for the majority of the spectra.

There remained twelve CENSORS sources for which there was significant
continuum detected, but insufficient signal-to-noise to assign a redshift.
For these sources the sky subtraction was repeated, instead using only
those sky fibres identified to be located in a similar region of the sky
as the target. This target-specific reduction was a lengthy procedure, but 
did improve the sky subtraction and yielded two further redshifts. 

\subsection{Long Slit Spectroscopy}

The data reduction technique was identical for all observations and
proceeded, using the \textsc{IRAF} data reduction package, as follows.
The raw 2D images were bias subtracted and flat fielded.  The sky
background was fitted and subtracted and the spectrum was extracted.  Flux
calibration information was extracted from a flux standard and applied to
each target, and wavelength calibration was performed on the basis of an
arc lamp observation.  The flux standards used varied for different
observing runs and include: Hiltner 600, Feige 56, LTT2415 and GD108.  The
spectra were extracted with apertures of width in the range 0.9 - 2.4
arcsec; the smaller extraction apertures were used for those sources with
very little continuum for which emission lines were the only features.
Subsequently, residuals from the subtraction of strong sky lines were
replaced by an interpolation of the surrounding continuum in order to ease
interpretation of the spectra.

\section{Spectroscopic Redshifts}
\label{results}

The method for identifying redshifts followed these steps:

\begin{itemize}
\item{If there were multiple emission lines, the ratios of their
wavelengths were used to identify them and derive an approximate redshift.
}
\item{The spectrum was then analysed in an \textsc{IDL} script which
measured the properties of all emission lines (and provided a more
accurate measurement of the redshift).  The properties measured for each
emission line were: the full width at half-maximum, which may be used as a
diagnostic for finding quasars; the emission line flux (for calibrated
spectra); the equivalent width.  The line properties are given for the
rest-frame in Table \ref{restable}. Note that because the Ly\,$\alpha$
line can have a distorted shape due to intervening matter absorption,
giving rise to an incorrect redshift estimate, this line is excluded from
the redshift determination.}
\item{Where there was only a single emission line, this line was generally
assumed to be one of the typically bright emission lines,
e.g. Ly\,$\alpha$, MgII or [OII], as these are the only ones which one
might expect to find alone.  MgII was easily distinguished by its broad
shape. Ly\,$\alpha$ and [OII] were distinguished by considering the
$K$--band magnitude and the redshift expected from the $K$--$z$ relation
for radio galaxies.}
\item{Where there were no emission lines, but there were absorption
features, the positions were measured in \textsc{IRAF}, using a Gaussian
fit, in order to establish the redshift.}
\item{In many spectra which did not lead to a redshift measurement via
emission lines or absorption features, continuum emission was nevertheless
detected. In some cases it was sufficiently strong that cross-correlation
with a template galaxy spectrum was attempted to see if a redshift might
be recovered. This was tried for three sources (CENSORS 34, 128 and 148),
using the elliptical galaxy template of \cite{kinney96} in the `fxcor'
package of \textsc{IRAF}. Since all of these candidates have 2df spectra
which have not been flux calibrated, both the galaxy spectra and the
template were divided by their smoothed spectra to improve the cross
correlation of features only. However, no redshifts were recoverable.}
\end{itemize}

The results of these observations are presented in Table \ref{restable}.
In order to have all the CENSORS spectroscopic information in one place,
this table presents the status of all CENSORS sources, including those
which have not yet been targetted and also those for which no redshift was
measured but an estimate has been made (see Section \ref{kzsection}).  A
column identifying which sources are believed to be starbursts or quasars
is also included (again, see Section \ref{kzsection}).  Where spectra were
successfully used to find a redshift, they are plotted in Appendix~C.
Uncertainties in the assigned redshift are also noted in Table
\ref{restable} and justification of the choice of redshift (where
necessary) is discussed source by source in Appendix \ref{indivnotes}.

It is finally noted that CENSORS 58 is located very close to a bright
galaxy (see Paper 2), prohibiting identification and spectroscopic
follow-up.  This source is removed from the sample for the analysis
presented in the remainder of this paper.  Since this is a random
selection, it is not believed that removing CENSORS 58 from the sample
will systematically affect the results.

\section{Redshift estimation via the ${\bf K}$--${\bf z}$ relation for
  radio galaxies} 
\label{kzsection}

In order to use this sample to investigate the evolution of the radio
luminosity function, a complete list of flux densities and redshifts is
required.  Paper 1 presents \14 flux densities for CENSORS and the current
work presents the spectroscopic redshifts.  In this section, redshift
estimates are described for the CENSORS sources which remain without a
spectroscopic redshift.

These redshift estimates were made via the $K$--$z$ relation for radio
galaxies.  $K$--band observations of radio galaxies are dominated by
emission from the old stellar population.  Since near-IR emission does not
change as rapidly with evolution of the stellar population as the shorter
wavelength emission, and radio galaxy hosts form a fairly homogeneous
sample of objects, a tight relation exists for radio galaxies between
$K$--band magnitude and redshift (\citeauthor{LL84} 1984,
\citeauthor{EalRaw96} 1996, \citeauthor{BLR98} 1998).

The $K$--$z$ relation for radio galaxies is known to vary slightly between
samples of different flux-density limits.  This has been shown by a number
of studies comparing the 3CRR, 6C and 7C radio surveys
(e.g. \citeauthor{EalRaw96} 1996, \citeauthor{Jarvis01KZ} 2001,
\citeauthor{Inskip02} 2002 and \citeauthor{Willott03KZ} 2003).  In Paper 2
it was shown that the CENSORS sample obeys a $K$--$z$ relation similar to
that of the 7C (7C is the faintest radio survey of those listed above and
CENSORS is fainter again) and this relationship has therefore been used to
estimate redshifts for the radio galaxies in the CENSORS sample.

\subsection{Identifying Starburst galaxies and Quasars in CENSORS}

The $K$--$z$ relation is applicable only to radio galaxies (for which the
$K$--band emission is dominated by an old elliptical galaxy). Therefore it
is necessary to identify any starburst galaxies or quasars in the CENSORS
sample before it can be used for redshift estimation.

\subsubsection{Starburst galaxies}
\label{sbgals}

In very deep radio surveys, star-forming galaxies begin to dominate the
radio source population (e.g. \citeauthor{Seymour04} 2004).
\cite{Jackson99} investigate a dual population scheme for AGN radio
sources using the two radio morphologies, FRI and FRIIs, as their basis.
They predict that around 10\% of radio sources at the flux density limit
of the CENSORS survey may be starburst galaxies. Therefore CENSORS may
contain a significant number of these sources.

A list of candidate starburst galaxies was produced from the CENSORS
sample. All sources with spectroscopic redshift greater than 0.5 or radio
luminosities greater than $2 \times 10^{24}$W\,Hz$^{-1}$ were removed
since the starburst population only dominates the radio galaxies at low
luminosities ($2 \times 10^{24}$ was chosen based on the local luminosity
functions for AGN and star forming galaxies, as being where the space
density of star forming galaxies is about two orders of magnitude below
that of AGN; \citeauthor{Sadler02} 2002). Of the remaining sources, those
with $K > 16.5$ were removed as these are highly likely to be at redshifts
greater than 0.5 (according to the $K$--$z$ relationship; described in
Section \ref{kzeqnsec}) plus any sources with jets in their radio
structure.  This left nine candidate star forming galaxies.

A useful way to distinguish star forming galaxies from AGN is to use a
diagram which compares the line flux ratios of \rm [OIII] 5007/\rm
H$\beta$~ and \rm [NII]6583/\rm H$\alpha$~(hereafter referred to as a BPT
diagram; \citeauthor{BPT81} 1981; \citeauthor{kauff03} 2003) because on
this plot AGN and star-forming galaxies fall in different regions.  Three
sources can be investigated in this way.

CENSORS 18 has $\log_{10} \left([\rm OIII] 5007/\rm H\beta\right) =
-0.281$ and $\log_{10} \left(\rm [NII]6583/\rm H\alpha\right) = 0.254$,
whereas for CENSORS 149 $\log_{10} \left(\rm [OIII] 5007/\rm H\beta\right)
= -0.237$ and $\log_{10} \left(\rm [NII]6583/\rm H\alpha\right) = -0.595$.
Thus CENSORS 149 falls on the star formation track of Fig. 1 of
\cite{kauff03} and is a star-forming galaxy, whereas CENSORS 18 lies in
the LINER region. CENSORS 18 also has broad lines in its spectrum,
confirming its AGN classification.  CENSORS 95 has both \Ha~~and
\nii~~detected in its spectrum, though no \oiii~~ or \Hb~~(suggesting it
is may be a heavily dust reddened source).  However the ratio $\log_{10}
(\rm [NII]6583/\rm H\alpha) = -0.602$, so this source is clearly more
consistent with being a star-forming galaxy.

Four more starburst candidates are CENSORS 93, 108, 140 and 148.  None of
these sources show the appropriate emission lines in their spectra to use
the BPT diagram.  However, it is possible to classify them on the basis of
their lack of an \oii\ detection.

\begin{figure}
\centerline{\psfig{file=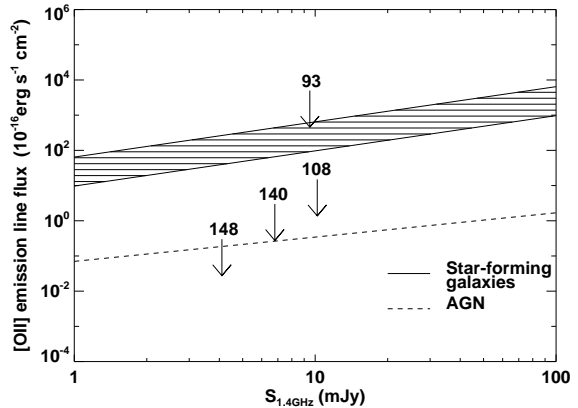,angle=90,width=8cm}}
\caption[]{The relationship between \oii\ line flux and radio luminosity
for star-forming galaxies and AGN, for a redshift of 0.24. This redshift
is chosen as it is similar to those of CENSORS 108 and 140, however the
difference for the redshifts of the other sources is not significant. The
shaded region shows the effect of 2 orders of magnitude of dust extinction
on the star-forming galaxies. The dashed line plots the expected relation
for AGN.  The limits for CENSORS 108, 140, 148 clearly indicate that they
are AGN. Although the limit for CENSORS 93 is consistent with either a
star-forming galaxy or an AGN, its spectrum is very typical of an old, red
elliptical.
\label{star_oii_plot}
}
\end{figure}

\cite{Bar97} provide a relation between the \oii\ line luminosity and the
(upper limit to the) star formation rate.  \citeauthor{condon90} (1990)
provide a relation between the radio luminosity and star formation rate of
star-forming galaxies.  Combining these results the relation between
\oii\ line emission and radio luminosity (assuming star formation as the
origin) is (cf. \citeauthor{best02_MS1054} 2002):
\begin{equation}
S_{\14} [\mu Jy] \sim 18.4 \left( \frac{f_{\rm [OII]}}{10^{-16} ergs^{-1}cm^{-2}}\right)  (1 + z)^{-0.8},
\end{equation}
where $f_{\rm [OII]}$ is the \oii ~line flux. This is plotted in
Fig. \ref{star_oii_plot} for a source at $z = 0.24$ (similar to the
redshifts of CENSORS 108 and 140, though the variation is not significant
for the other sources).

The \14 flux for a given line luminosity that is expected for AGN may also
be estimated. \cite{willott99} provide a relation between 151\,MHz radio
luminosity and \oii ~line luminosity for radio-loud AGN. Following
\cite{best02_MS1054}, this can be scaled to \14 assuming a spectral index
of $\alpha = 0.8$, and extrapolated to low flux densities, giving:
\begin{equation}
S_{1.4\,\rm{GHz}} [\mu Jy] \sim 4.7\times 10^{4}\left(\frac{f_{\rm
[OII]}}{10^{-16} ergs^{-1}cm^{-2}}\right)^{1.45}.
\end{equation} 

\noindent Thus, the upper limit for emission line flux for these four
candidates can be compared to the radio luminosity to investigate its
origin. Although the 2dF spectra for CENSORS 93, 108, 140 and 148 are not
flux calibrated, a limit on the line flux may be estimated by combining
the $V$--band continuum flux density (from the EIS broad band magnitude)
with the upper limit to the line equivalent widths. The $V$--band is close
enough in wavelength to the redshifted line for this to be reasonable.

 Fig. \ref{star_oii_plot} plots the limiting \oii\ line fluxes against
their radio fluxes and the theoretical relations for star-forming galaxies
and AGN.  The shaded region shows the effect of 2 magnitudes of dust
extinction on the star-forming galaxies.  The limits for CENSORS 108, 140
and 148 are clearly closer to the dashed AGN line.  The limit for CENSORS
93 is consistent with the star-forming region of the plot.  However, the
spectrum of CENSORS 93 in Figure \ref{spec_plots}, shows that it is
typical of an old, red elliptical galaxy and not a star-forming galaxy.
Therefore all of these candidates are considered to be AGN.

The final two star-forming candidates, CENSORS 124 and 146, are both very
low redshift. Their radio flux densities in the VLA BnA array observations
are very much lower than those in the NVSS, indicating that they are
diffuse sources.  Such diffuse radio emission is more often associated
with starbursts and not AGN.  In addition the radio luminosities of these
sources are $\simeq$~~$10^{22}$\whz1 at \14, making them more likely to be
star-forming galaxies rather than AGN.

The final list of radio sources due to star formation in the CENSORS
sample is: CENSORS 95, 124, 146, 149.

\subsubsection{Quasars}

The SExtractor parameter $S/G$ is provided by the EIS catalogue, and
varies from 0.0 to 1.0, 1.0 being the most point-like.  In Paper 1, 11
sources were identified as possible quasars on the basis of their
stellaricity measure being $S/G > 0.9$.  Some sources with a slightly
lower stellaricity were also included as possible quasars on account of
their blue colour ($S/G > 0.6$ and $B-I<1$). From the spectra presented in
this work, quasars may be identified by having broad emission lines
(broader than the $\simeq$~~1000\kms1 line width of radio
galaxies). Sources are considered candidates for quasars if they have
permitted lines, other than \lya\ (which can be broadened by resonant
scattering), which are significantly broader than 1000\kms1.

Combining these two methods, CENSORS 6, 7, 10, 11, 29, 39, 44, 48, 92 and
116 are classified as quasars by both methods, and are considered certain
quasars. CENSORS 63, 82, 100, 126, 129 are classified as quasars from the
imaging data, but do not show the broad emission lines, whilst CENSORS 18,
37, 38, 49, 52, 92, 114, 129, 135 are spectroscopic quasar candidates but
were not classified as quasars from the imaging data. These additional
quasar candidates were considered individually (as described in the notes
on individual sources which are presented in Appendix \ref{indivnotes}),
and CENSORS 37, 38, 114 and 135 were added to the list of quasars.

In total, the quasars within the CENSORS sample are CENSORS 6, 7, 10, 11,
29, 39, 44, 48, 63, 82, 92, 100, 116, 126.

\subsection{Estimating redshifts using the ${\bf K}$--${\bf z}$ relationship for radio galaxies}
\label{kzeqnsec}
In Paper 2, it was shown that the $K$-$z$ relationship, based upon the 7C
radio selected sample from \cite{Willott03KZ} is applicable to CENSORS.
In order to estimate the redshift from the $K$-band magnitude, the fit of
$K$ as a function of $z$ (as the $K$-$z$ relationship is traditionally
expressed), should not simply be mathematically inverted as that leads to
over-predictions of the redshifts of faint K-sources (due to the small
number of high redshift sources). Rather, the relation should be re-fitted
using K as the independent variable. This is discussed in more detail in
\cite{cruz07}. 

A second order polynomial, $\log_{10}z(K)$, was produced by fitting to the
7C data via a least squares method, providing the following equation:

\begin{equation}
\label{kzeqn}
\log_{10}z = 0.003K^2 + 0.11K - 2.74
\end{equation}

\noindent The scatter, $\sigma_{7C}$, about the predicted $\log_{10} z$
for the 7C sources is 0.14.  

This relation is applied to the $K$--band magnitudes measured for those
CENSORS sources without spectroscopic redshifts in order to derive a
redshift estimate; the redshift estimate is then used to correct the
photometry to the standard aperture used to define the $K$--$z$
relationship (see Paper 2 for details) and the process is repeated
iteratively.  The aperture chosen for this calculation was the one which,
by eye whilst executing the photometric measurements, covered the most
source emission without being unnecessarily large (in order to keep
photometric errors as low as possible).  Where no detection was made in
the $K$--band, the 2-$\sigma$ limiting magnitude, based on a 1\arcsec\
radius aperture, was used to derive a redshift estimate as described
above. This estimate was then reduced to a 95\% confidence lower limit,
based upon the scatter in $\log_{10} z$ in the $K$--$z$ relation.  The
results of the redshift estimation are presented in Table
\ref{spec_obs_details}.

\subsection{$I$--$z$ relationship for CENSORS 112}

The FORS1 spectrum of CENSORS 112 has faint blue continuum, but no
features. A $K$--band magnitude is not available for this source, so it is
not possible to use the $K$--$z$ relationship to estimate its redshift. A
redshift may be estimated using the $I$--$z$ relationship, following
\cite{CENSORS1}.  Using this relation the redshift for CENSORS 112 is
estimated to be $z = 1.75$.

\begin{figure}
\begin{tabular}{c}
\centerline{\psfig{file=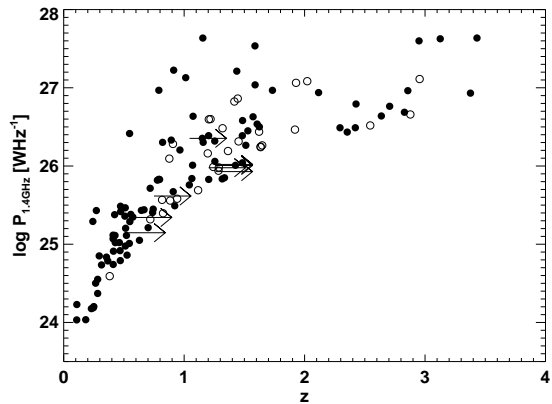,angle=90,width=8cm}}\\
(a)\\
\centerline{\psfig{file=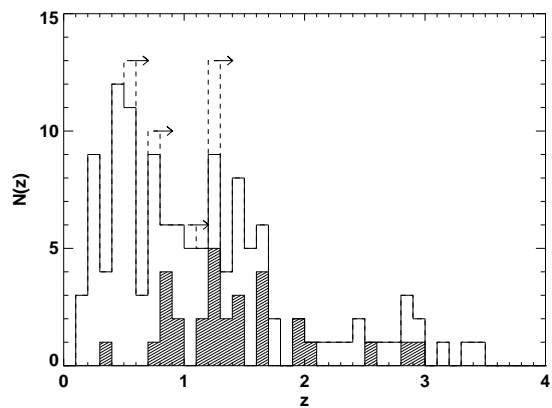,angle=90,width=8cm}}\\
(b)\\
\end{tabular}
\caption[] { (a) The coverage of the radio-luminosity--redshift plane
provided by the CENSORS sample. Closed circles indicate spectroscopic
redshifts (including those with single line redshift assignments) and open
circles indicate estimated redshifts on the basis of the $K$-- or $I$--$z$
relations. Arrows indicate 95\% confidence, lower limits to redshifts for
sources with no spectroscopic redshift and no $K$--band detection as
described in the text.  Note that if these sources were at higher
redshifts, their luminosities would also increase.  (b) The redshift
distribution for the CENSORS sample. The hashed area represents those
sources with redshifts estimated from the $K$--$z$ relation for radio
galaxies. The unfilled region represents spectroscopic redshifts and the
dashed/arrowed region represents lower limits to redshifts based upon the
$K$--$z$ relation.
\label{pzplane}
}
\end{figure}

\section{The CENSORS redshift distribution}
\label{NZ}

\subsection{A summary of the current redshift status of the CENSORS sources}

Excluding CENSORS 58 there are 149 sources, of which 143 have been
spectroscopically targetted.  Of the 6 sources not targetted, 2 have
redshifts available from NED and the remaining 4 were not observed due to
poor weather or lack of observing time.  The sample as a whole is 71\%
spectroscopically complete and this figure does not change if only the 136
sources complete to a flux-density of 7.2\,mJy at 1.4\,GHz are considered.
A break down of the relative certainties of the redshifts ascribed is
given in Table \ref{specsum_table}.

\begin{table}
\begin{center}
\begin{tabular}{|c|c|}
\hline
Total Sources 				&149	\\\hline
Total targetted				&143	\\\hline
Literature redshifts			&2	\\\hline
Total secure redshifts$^{\dagger}$	&81	\\\hline
Less secure redshifts$^{\dagger}$	&2	\\\hline
Well justified single line redshifts	&11	\\\hline
Less secure single line redshifts	&9	\\\hline
\end{tabular}
\caption{A summary of the results of spectroscopic observations of CENSORS. $\dagger$ Not including single line redshifts.\label{specsum_table}}
\end{center}
\end{table}

Of those sources without spectroscopic redshifts, one source has a
redshift estimated from the $I$-$z$ relation, and all but 9 more have
redshifts estimated from the $K$-$z$ relation. For the remainder, a
non-detection in the $K$ band offers a lower limit to the redshift.  Table
A1 lists the final redshift assigned to each source and a comment on the
origin of the redshift estimate.  This table includes all spectroscopic
and estimated redshifts for CENSORS. In the column listing the redshifts,
the symbols `{\bf ?}' and `{\bf s}' are used to clearly identify those
redshifts which are considered to be less secure and/or derived on the
basis of a single line.

On the basis of the presented spectroscopic and estimated redshifts, the
radio luminosity--redshift plane and redshift distribution are plotted for
the CENSORS sample in Fig. \ref{pzplane}.  This figure and, henceforth,
the discussion, is limited to the 134 sources which are complete to
7.2\,mJy at 1.4\,GHz and whose radio emission is not due to star
formation.  As expected, the spectroscopic completeness of the sample is
greater at low redshifts, say $z \lesssim 1.1$ at which it is
$\simeq$~~82\%, compared with higher redshifts at which it is reduced to
$\simeq$~~56\% (assuming the redshift estimates are roughly correct).
This is to be expected as it is the sources likely to be at high redshifts
that are more likely to be without a detection of the host galaxy.  In
addition, measuring redshifts of galaxies which lie between of 1.1 and 2.2
(the redshift desert) is difficult as neither the [OII] or \lya~lines are
present in the spectrum.

\subsection{Comparison with the Dunlop and Peacock models}

The development of the CENSORS sample was primarily motivated by the need
to provide increased coverage of the luminosity-redshift plane to improve
our knowledge of the cosmological evolution of the radio luminosity
function.  Now that the required redshift information has been assembled,
a preliminary assessment of the potential impact of this sample on the
determination of the evolving RLF can be undertaken.  One way to do this
is to compare the redshift distribution of the CENSORS sample with that
predicted by existing models of the RLF.  The most appropriate comparison
is with the best-fit models of \citet{DP90}. Not only is this work the
most complete of its kind, but it is also based upon high frequency data
(2.7\,GHz) and may be used to make predictions for the CENSORS sample with
the least errors introduced by making the correction to 1.4\,GHz.

In \cite{DP90}, five free-form models were fitted to the data available at
the time.  Free form model 1 (FF1) describes the number density as a
series expansion of 0.1($\log P - 20$) and 0.1$z$. FF2, 3, 4, and 5 are
similar but explore the effect of imposing cut-offs at high luminosity
(FF2) or redshift (FF5), changing the nature of the coordinates in
luminosity and redshift (FF3), and terminating the integration of the
luminosity-redshift plane at $z=5$ instead of $z=10$ (FF4).  In addition
two models with more general assumptions regarding the evolution were also
fitted.  These were pure luminosity evolution and luminosity--density
evolution.  \cite{DP90} model the flat and steep spectrum radio sources
separately. Since these populations cannot yet be separated within the
CENSORS sample, it is the combined steep and flat spectrum predictions
from \cite{DP90} which are used in the comparison.

Table \ref{kstest} provides the probabilities, based upon the K--S tests,
that a given model could be produced by the same source distribution as
the data. Here `Data 1' is used to describe the best estimated redshift
distribution.  This is essentially the distribution plotted in Figure
\ref{pzplane}(b) however the sources for which redshifts are based upon
$K$--band limits are placed at their best redshift estimate, as calculated
by Equation \ref{kzeqn}, as opposed to the 95\% confidence lower limit to
their redshift.

\begin{table}
\begin{center}
\begin{tabular}{|c|c|c|c|}
\hline
MODEL&	N$^o$ Sources &	Data 1	&Data 2 \\
\hline
1&   122 &	0.35	&0.61	\\	
2&   122 &	1$\times$10$^{-6}$	&2$\times$10$^{-4}$	\\	
3&   148 &	0.005	&9$\times$10$^{-4}$	\\
4&   141 &	0.16	&0.09	\\
5&   139 &	0.07	&0.15	\\
PLE& 131 &	3$\times$10$^{-6}$	&1$\times$10$^{-4}$	\\
LDE& 115 &	2$\times$10$^{-5}$	&7$\times$10$^{-4}$	\\
\hline
\end{tabular}
\caption{The number of sources predicted to the CENSORS flux density limit
(compared to 134 observed) for each of the DP90 models, and the
probabilities, according to a K-S test, that the redshift distribution of
those sources could be drawn from the same source distribution as the
CENSORS data. Here `Data 1' refers to the best estimated redshift
distribution in which all estimated redshifts are assumed correct. In
`Data 2' each source which has an estimated redshift has had the single
`best' value of estimated redshift (used in `Data 1'), replaced with a
spread of values which integrate to unity as described in the text. Note
that the K-S test probability gives only the probability that the
normalised distributions are the same, and does not address the
differences in the total numbers of sources predicted.\label{kstest}}
\end{center}
\end{table}

Models 1, 4 and 5 have a $>$5\% chance of matching the observed
(normalised) redshift distribution, with other models strongly ruled out.
However, model 1 badly under-predicts the number of sources at these faint
radio flux densities, and we are left to conclude that models 4 and 5
provide the best description of the data.  Interestingly, this was also
the finding of \cite{Waddington01}, although the actual agreement with the
redshift distribution of their sample was somewhat better than found here.

\begin{figure}
\begin{tabular}{c}
\centerline{\psfig{file=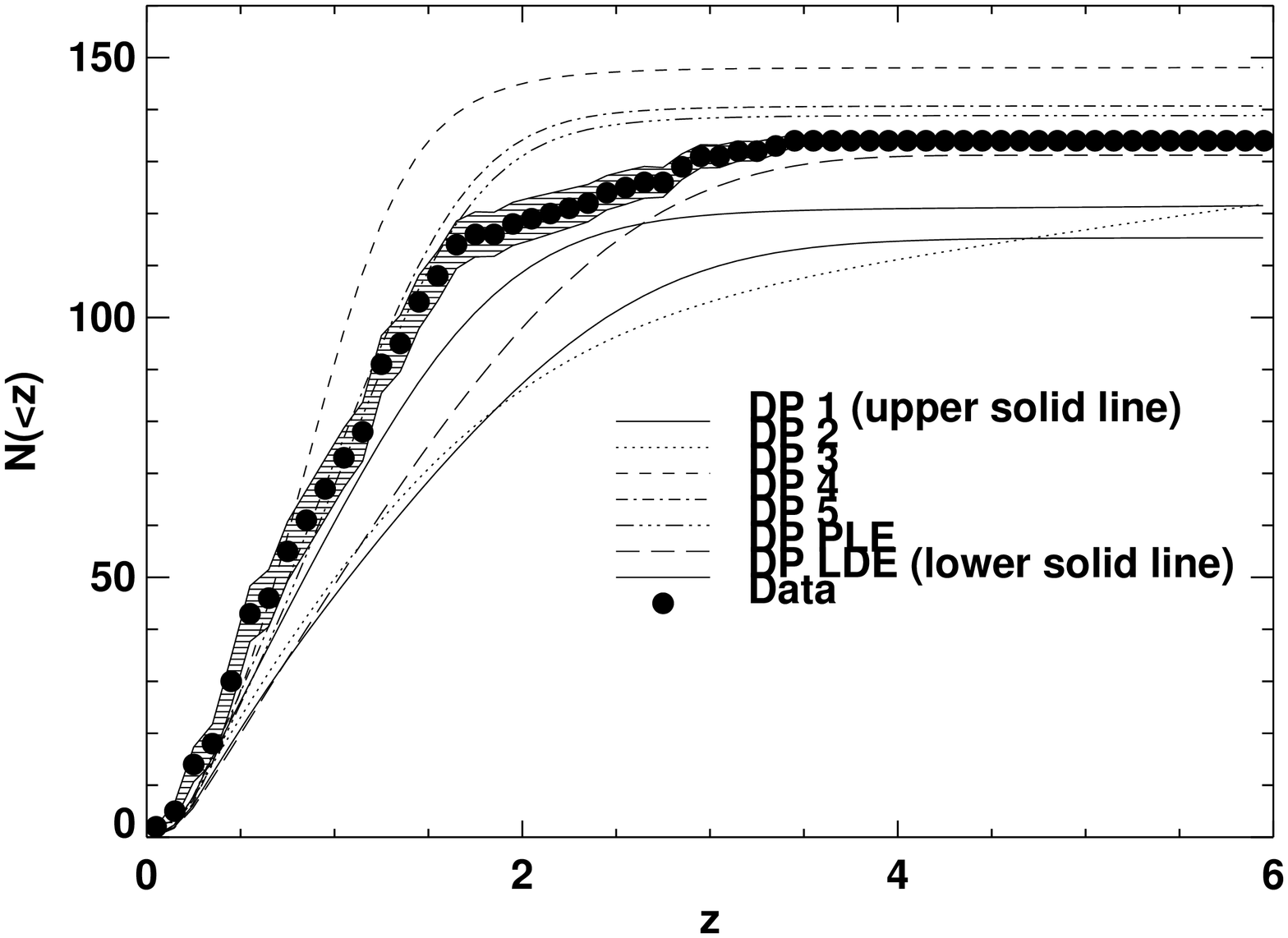,angle=90,width=8cm}}\\
{\bf (a)}\\
\centerline{\psfig{file=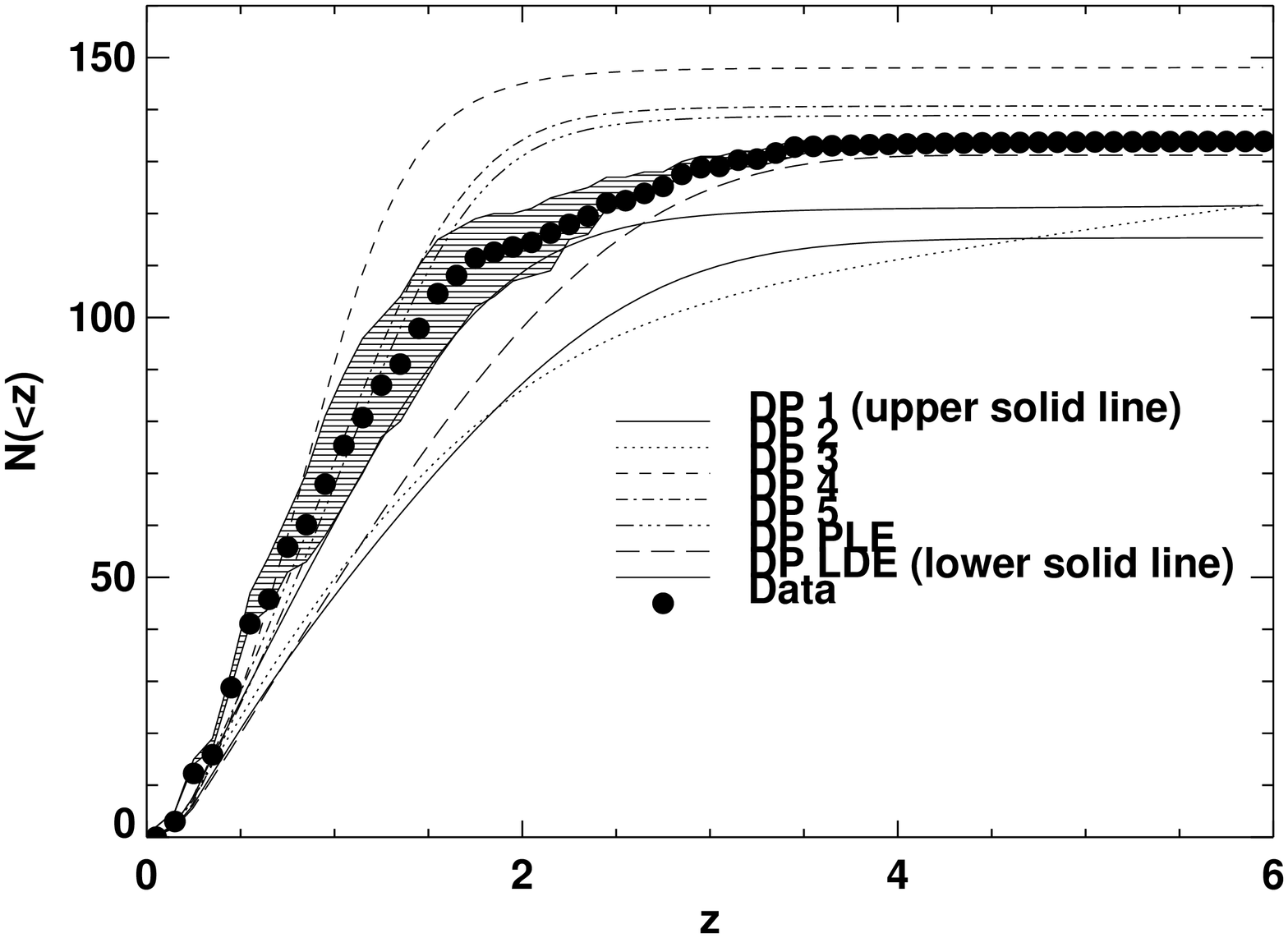,angle=90,width=8cm}}\\
{\bf (b)}\\
\centerline{\psfig{file=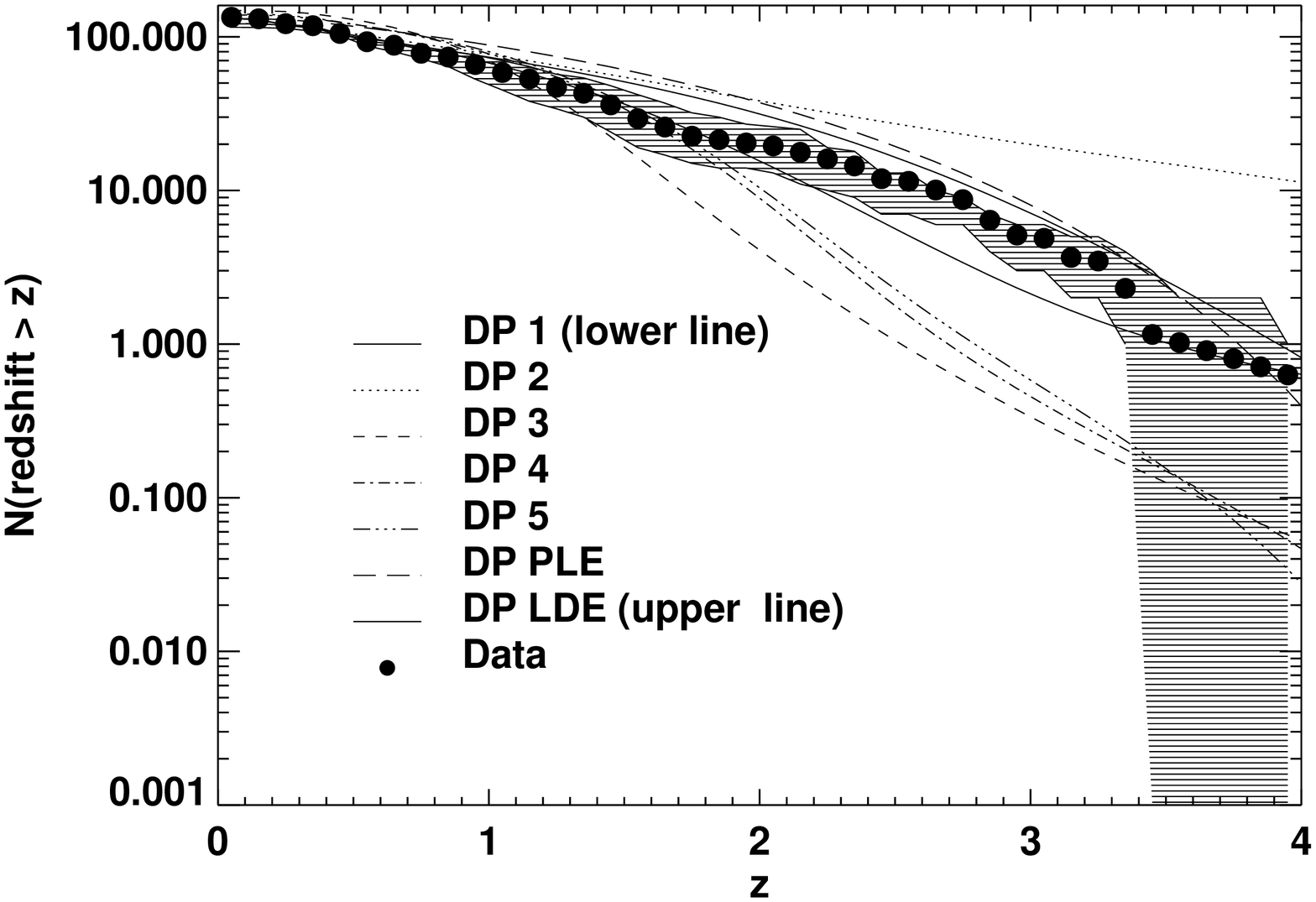,angle=90,width=8cm}}\\
{\bf (c)}\\
\end{tabular}
\caption{The cumulative redshift distributions, N($<z$), of the CENSORS
sample.The shaded regions indicate the uncertainty as calculated by finite
number statistics, in {\bf (a)}, and the uncertainty due to the
1--$\sigma$ spread in the $K$--$z$ relation about the estimated redshifts,
in {\bf (b)}. {\bf (c)} plots the redshift distribution, N($>z$), which
highlights the differences between the smoothed data and the models at
high redshifts.\label{cnz}}
\end{figure}

Figure \ref{cnz}(a) presents the comparison graphically and can be
usefully compared with Figure 9 from \cite{Waddington01}. It plots the
cumulative redshift distribution, N($<z$), of CENSORS (solid circles) as
compared with the best fitting models of \citet{DP90}.
The shaded region indicates the error as calculated by finite number
statistics.  It is clear that none of the \citet{DP90} models are a really
good fit to the data over all redshifts.  In this plot the distributions
are plotted in terms of the absolute number of sources, thus highlighting
where the models may have a reasonable shape but predict the wrong number
of sources (the K--S test assesses only the maximum difference between the
normalised distributions).  However this comparison is somewhat confused
by the assumption in this figure that the distribution for the data may be
represented by the estimated redshifts without account for the
uncertainty.

In order to account for the uncertainty in the distribution a smoothed
version (Data 2) has been produced.  In this smoothed distribution, the
$K$--$z$ redshifts are replaced by a Gaussian distribution about the best
estimated redshift of the width $\sigma = \sigma_{7C}$ (the width of the
scatter of the $K$--$z$ relation about $\log z$). The total contribution
is one source but spread over a range in redshifts.  In a similar way the
$K$--$z$ estimated redshifts that are based upon limiting $K$--band
magnitudes are replaced by a source which is spread $\sigma = \sigma_{7C}$
to lower redshifts and $\sigma = 2\sigma_{7C}$ to higher redshifts than
the best estimated redshift from Equation \ref{kzeqn}.  Note that adopting
a $2\sigma$ spread to high redshifts is somewhat arbitrary, but changing
this to $4\sigma$ did not affect the conclusions.  CENSORS 112, for which
the $I$--$z$ relation is used, has been treated in the same way as the
$K$--$z$ sources.

Figures \ref{cnz}(b) and (c) plot this smoothed distribution (filled
circles) in comparison to the \cite{DP90} models. The shaded regions show
the change if all K-band estimated redshifts are moved up or down by
1-sigma.  These figures show that the uncertainty in the distribution is
dominated by the scatter in the $K$--$z$ relation and that more
spectroscopic redshifts, or more accurate photometric redshift estimates,
are required in order to tie down the shape of the distribution precisely.
The fact that none of the \citet{DP90} models provides a very good fit to
the CENSORS data demonstrates that the sample is probing a region of
parameter space which was not available before.  CENSORS may therefore
play an important role in expanding on previous work.

\section{Summary}
\label{summary}

Of 150 CENSORS sources, 143 have been spectroscopically targetted using
the AAT, VLT, WHT and ESO 3.6m telescopes.  These observations, plus
results from the literature, have resulted in secure spectroscopic
redshifts for 56\% of the sample based upon two or more emission lines or
absorption features.  In addition 22 sources have less certain
redshifts. Of these 20 are based upon a single emission line, but 11 are
considered secure redshifts when photometry and the lack of other bright
lines are taken into account.  Including all of these redshifts, the
sample is 71\% spectroscopically complete.

Quasars and starbursts have been identified in the sample, thus allowing
redshift estimation, on the basis of the $K$--$z$ relation for radio
galaxies (derived from the 7C radio survey in Paper 2), for those sources
without a spectroscopic redshift.  For CENSORS 112, an estimate based on
an $I$--$z$ relation was made.

The resulting redshift distribution has been compared to the predictions
of the best fitting models from \citet{DP90}.  None of these models
provides a convincing match to the data.  These results demonstrate that
this new sample probes a region of radio luminosity and redshift space
which was not available at the time of the \citet{DP90} investigation.
Since that investigation has been the most complete to deal with all radio
sources selected at relatively high frequencies, it is clear that a new
study into the evolution of radio sources is timely.

\section*{Acknowledgements} 

MHB is grateful for the support of a PPARC research studentship, and JAP
is grateful for the support of a PPARC Senior Research Fellowship.  PNB
thanks the Royal Society for generous financial support through its
University Research Fellowship scheme.  Observations were made using the
Anglo-Australian Telescope, the William Herschel Telescope, the ESO Very
Large Telescope at the Paranal observatory (69.A-0047, 71.A-0622,
76.A-0745) and the ESO 3.6m telescope at La Silla observatory (70.A-0225)

\label{lastpage}
\bibliography{/Users/mbrookes/LaTeX/mhb} 
\bibliographystyle{/Users/mbrookes/LaTeX/mn2e} 

\appendix

\section{Details of the spectroscopic observations and results}

Table~A1 provides full details of the spectroscopic observations, the
resulting spectra and line properties, and the redshift estimates for
sources without spectroscopic redshifts.

\begin{table*}
\centerline{\psfig{file=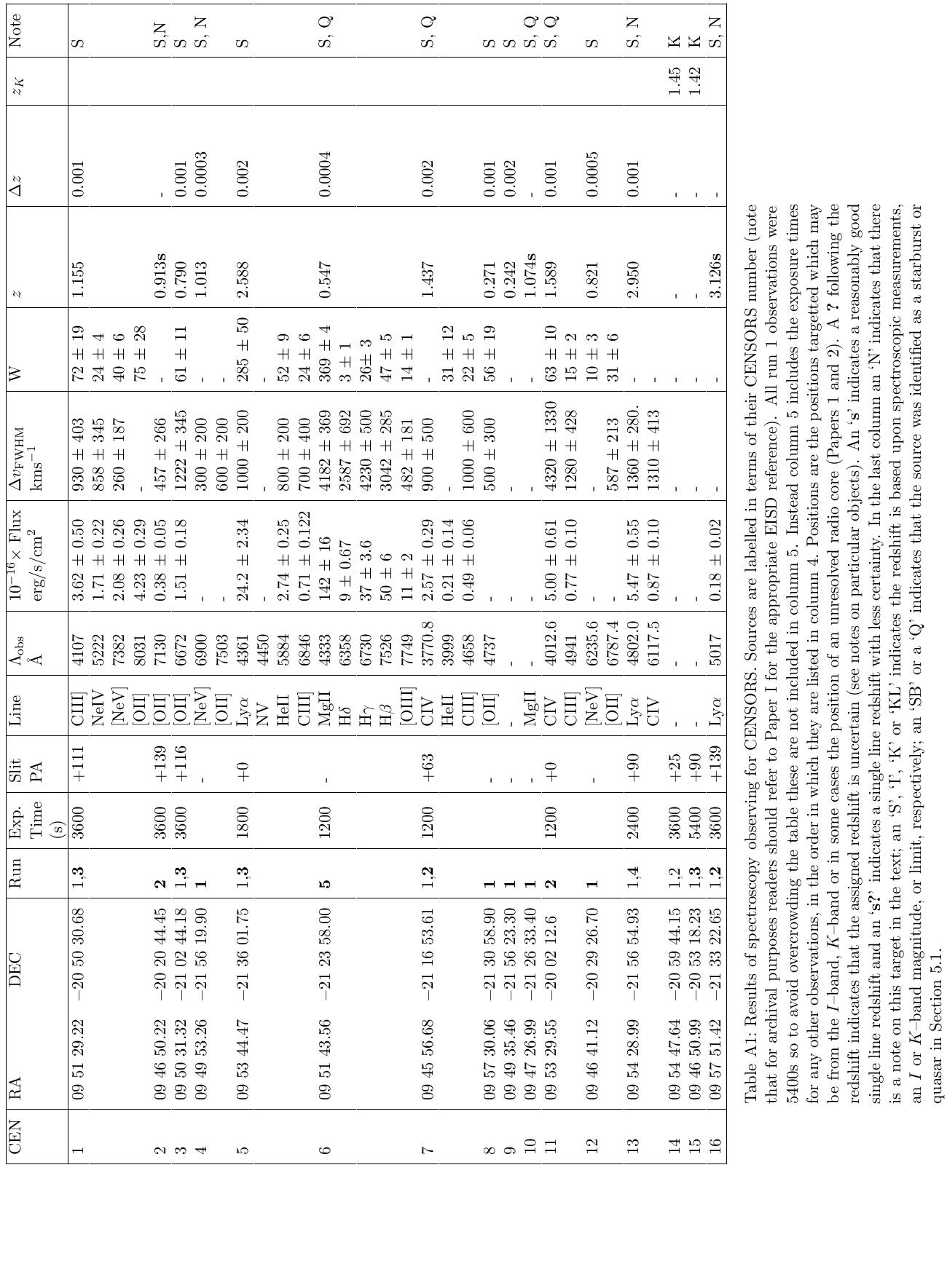,angle=0,width=\linewidth}}
\caption{\label{restable}}
\end{table*}

\begin{table*}
\centerline{\psfig{file=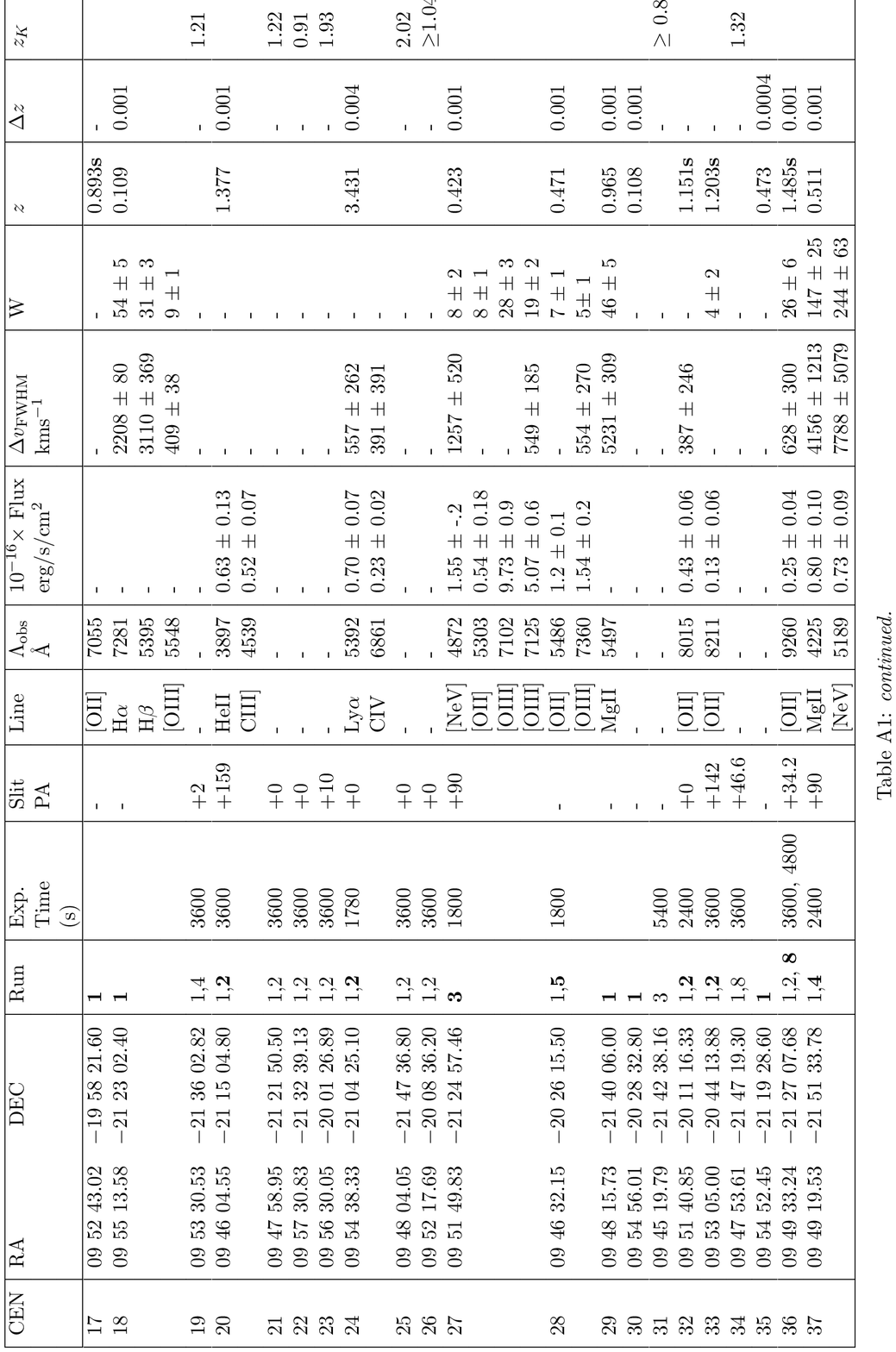,angle=0,width=\linewidth}}
\end{table*}

\begin{table*}
\centerline{\psfig{file=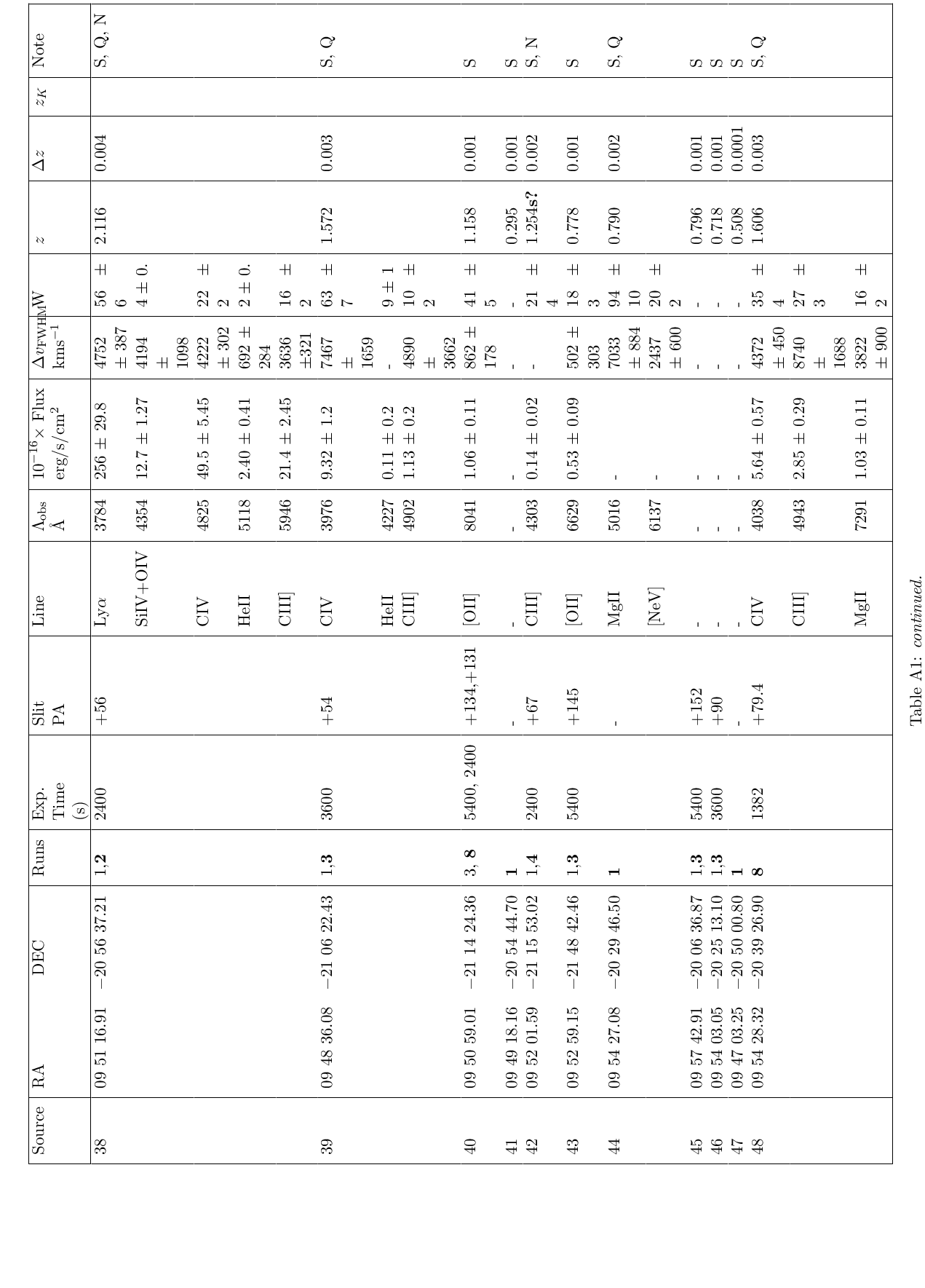,angle=0,width=\linewidth}}
\end{table*}

\begin{table*}
\centerline{\psfig{file=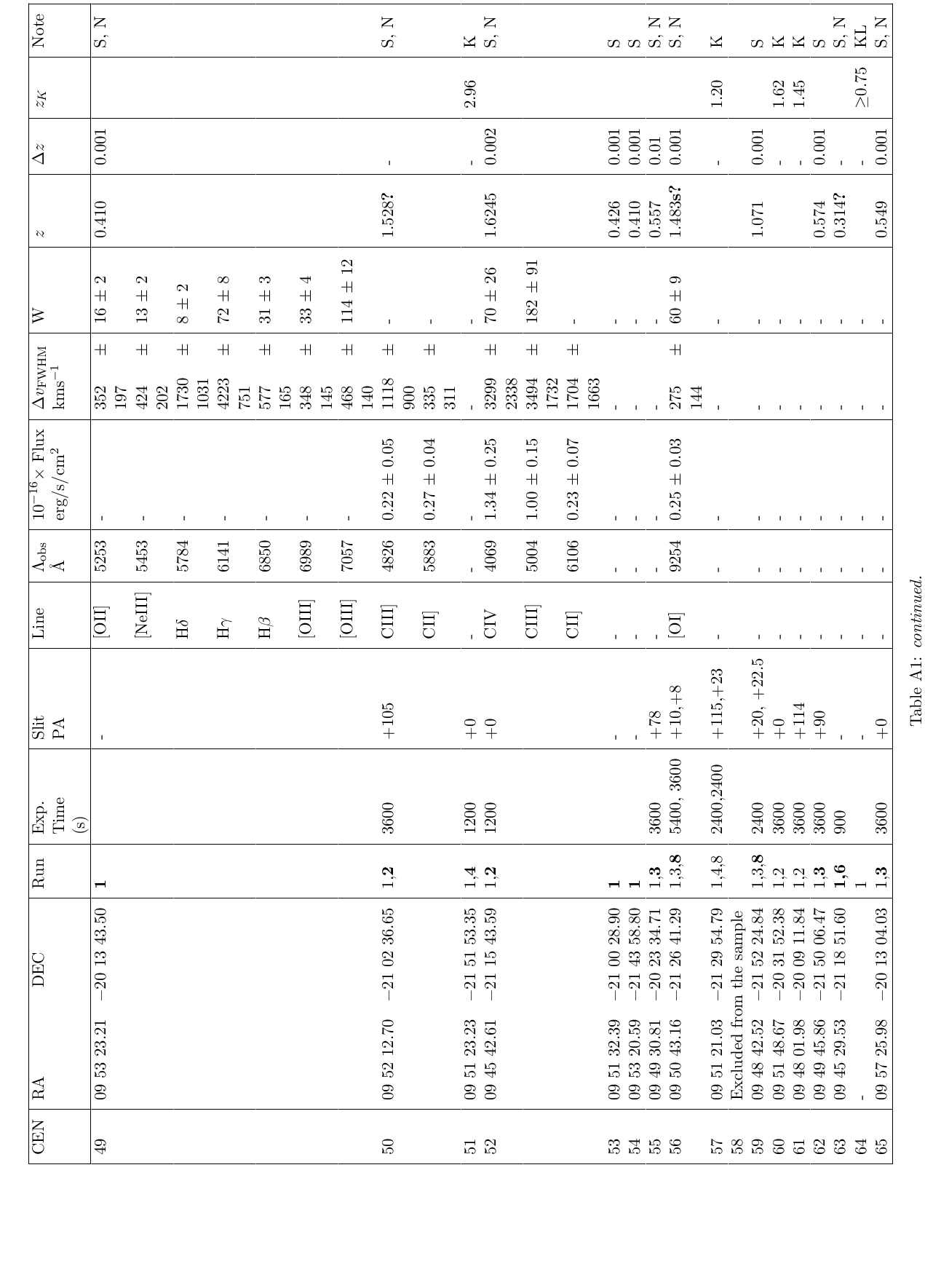,angle=0,width=\linewidth}}
\end{table*}

\begin{table*}
\centerline{\psfig{file=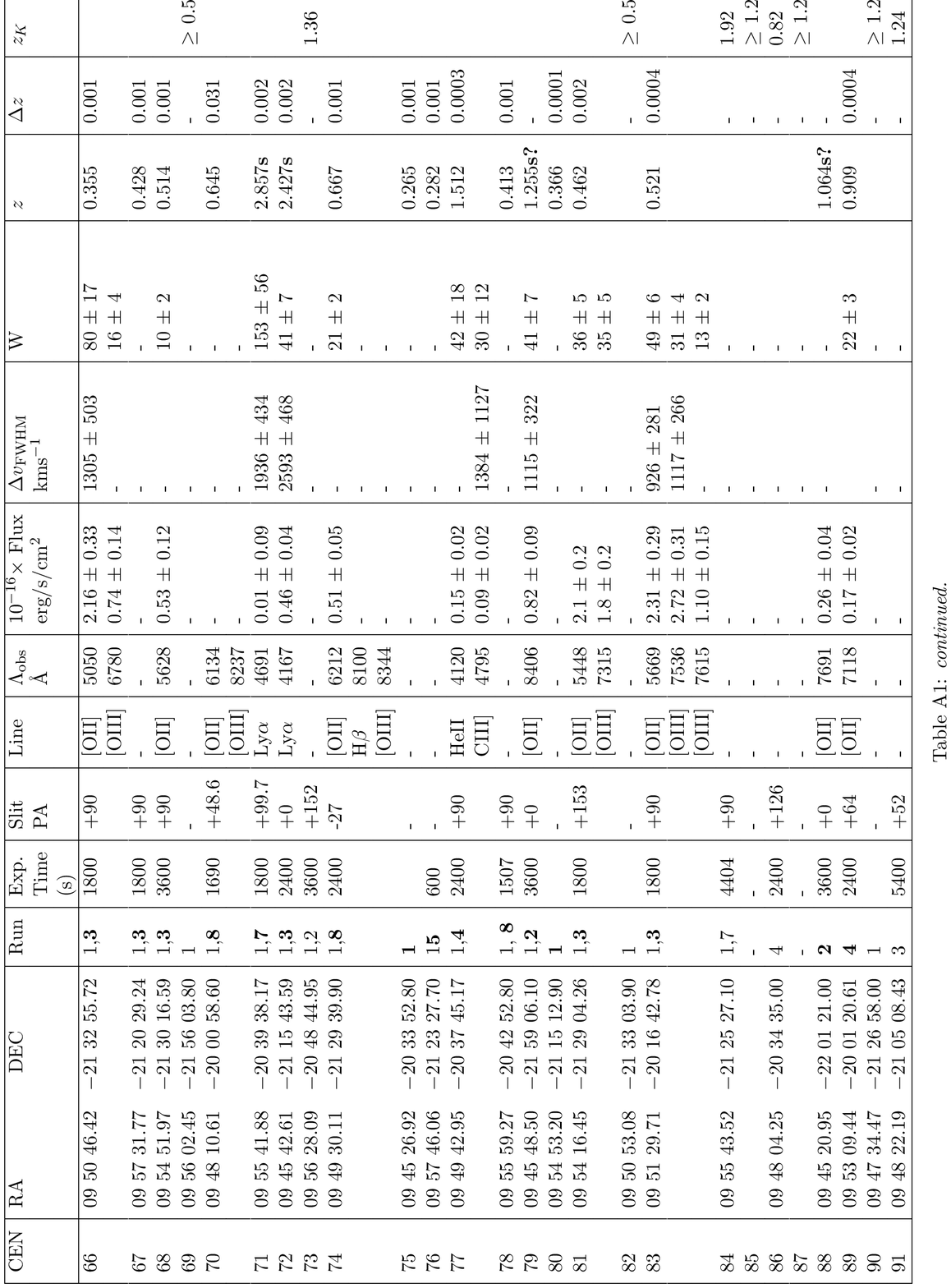,angle=0,width=\linewidth}}
\end{table*}

\begin{table*}
\centerline{\psfig{file=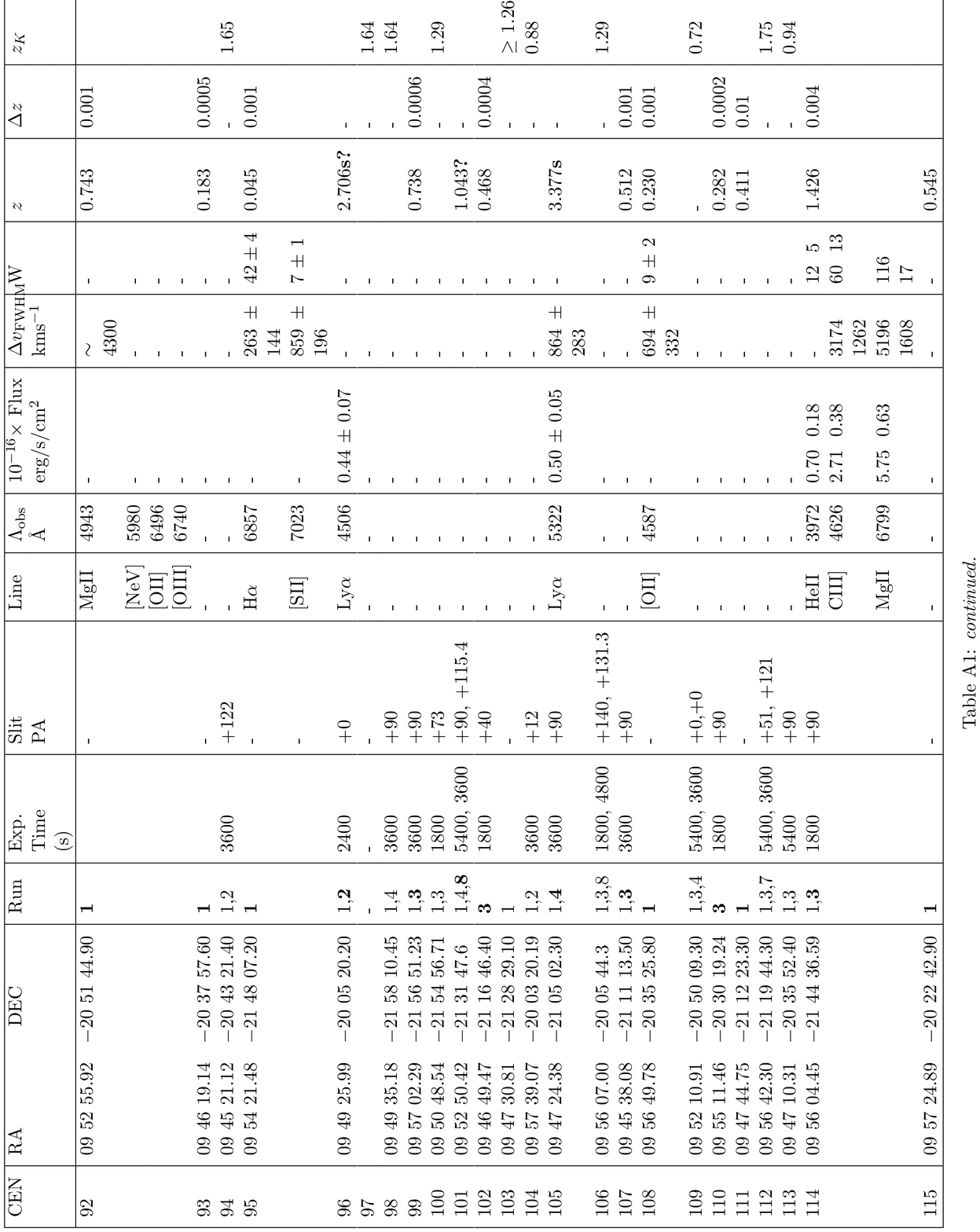,angle=0,width=\linewidth}}
\end{table*}

\begin{table*}
\centerline{\psfig{file=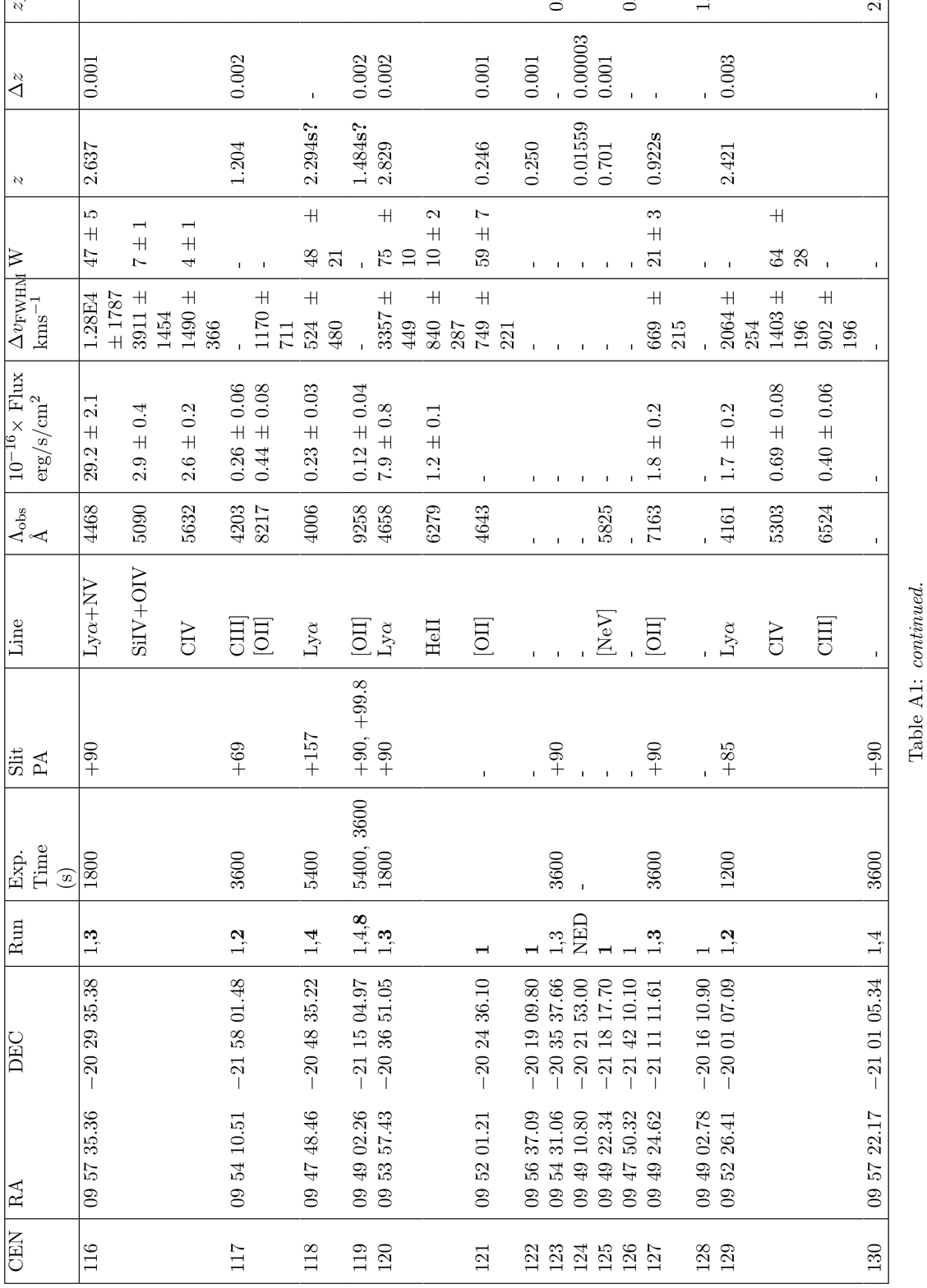,angle=0,width=\linewidth}}
\end{table*}

\begin{table*}
\centerline{\psfig{file=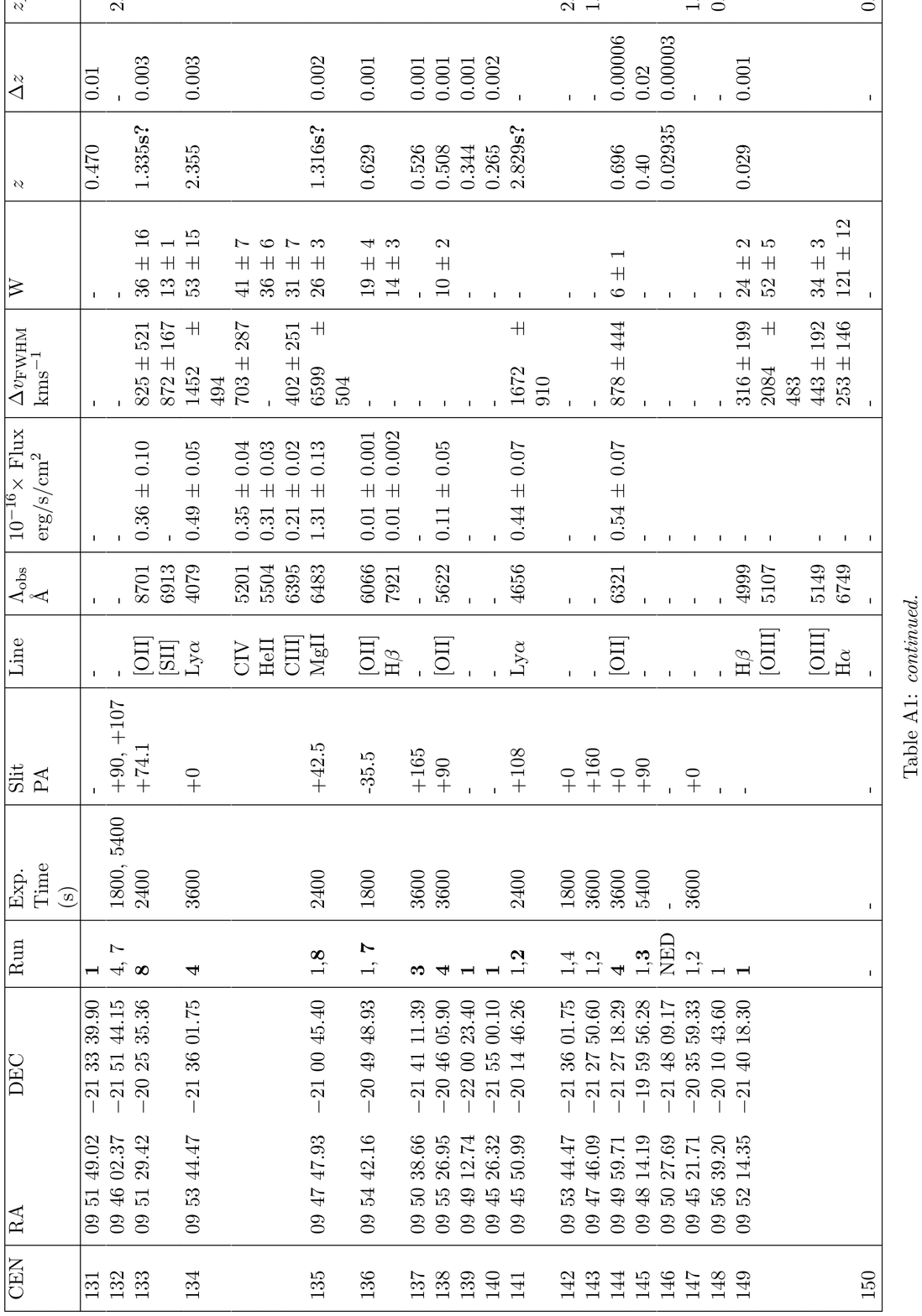,angle=0,width=\linewidth}}
\end{table*}

\section{Notes on individual sources}
\label{indivnotes}

This section provides details on the interpretation of the spectroscopic
observations in terms of both measured redshifts and whether sources are
identified as starforming galaxies or quasars.

\noindent 
{\bf CENSORS 2:} This is a single line redshift. The most likely
candidates for a single line are the bright emission lines [OII] and
Ly$\alpha$ which would place this source at $z$ = 0.91 and $z$ = 4.87
respectively. A $K$--band magnitude of 19.00(13) (where the error, in
units of 0.01 mags, is given in brackets) in a 1\arcsec aperture makes the
lower redshift more likely. Adding further weight to that conclusion are
two other galaxies within 6\arcsec and on either side, which also have
single emission lines at wavelengths within 100\AA~of the detected
line. It is therefore likely that these three galaxies form an interacting
system.

\noindent 
{\bf CENSORS 4:} In Paper 1 two likely candidates were listed for this
source. This AGN spectrum now identifies the correct host galaxy to be the
northerly of those two.

\begin{figure}
\centerline{\psfig{file=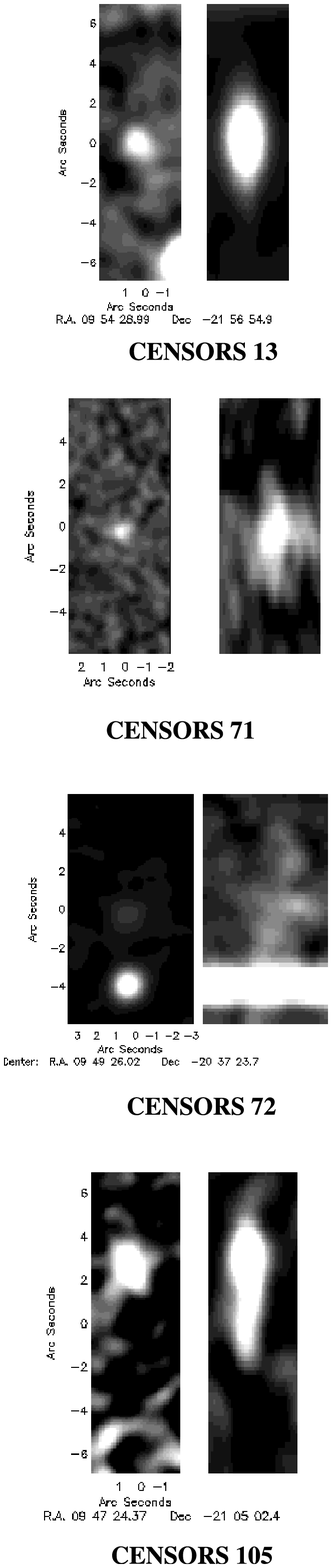,width=6cm}}
\caption[]{ Extended emission in CENSORS 13, 71, 72, 105 which is believed
to be \lya~emission.  The left panels show the $K$--band image and the
right hand panels show the spectrum, dispersed in the x--direction and
matched spatially in the y--direction.  Note that in the case of CENSORS
72 there is bright star beneath the faint host in this image, the emission
associated with the host is not a feature in the stellar spectrum.
\label{lya_fig}
}
\end {figure}

\noindent 
{\bf CENSORS 13:} This source has significantly extended Ly $\alpha$
emission as shown in Fig. \ref{lya_fig}.

\noindent 
{\bf CENSORS 16:} This single line spectrum has been assigned a redshift
corresponding to detection of Ly $\alpha$. With a 1\arcsec aperture
$K$--band magnitude of 19.85 this cannot be [OII] at $z$ = 0.35 according
to the $K$--$z$ relation. Although there is a hint of a He II (1640)
confirming line in the extracted spectrum this by no means certain from
the 2D image.

\noindent 
{\bf CENSORS 17:} Identifying this line as [OII] is consistent with its
18.07(20) $K$--band magnitude (1.5\arcs~ aperture).

\noindent
{\bf CENSORS 18:} This object has clear broad emission lines but is
identified with a low redshift, extended galaxy, and so is a broad line
radio galaxy (BLRG). It is not considered a quasar in this work.

\noindent
{\bf CENSORS 20:} No features were found at the centre of the target. The
radio structure is slightly double lobed and at the centre of the main
lobe two emission lines, but no continuum, were found. These have a
wavelength ratio of 1.164 which matches either CIII] and HeII or [OII] and
\Hb. The latter would give a redshift of 0.04 which does not match the
limiting $K$--band magnitude of 19.4.  Hence the former line
identifications are assumed.

\noindent 
{\bf CENSORS 31:} This source has two potential host galaxy candidates
based upon $I$ and $K$--band imaging. However both are misaligned to the
radio axis. The western candidate was identified as a star via 2df
spectroscopy and the eastern candidate was then targetted with FORS1. This
target is a low mass star as identified by a KI absorption feature (as
compared to examples in \citeauthor{leggett96} 1996). This implies that
the actual host galaxy must remain undetected to a limit of 18.2 in $K$.

\noindent 
{\bf CENSORS 32:} Identifying this line as [OII] is consistent with its
18.37(13) $K$--band magnitude (1\arcs aperture).

\noindent 
{\bf CENSORS 33:} This single line is convincing in the 2D image. The
19.17(12) K band magnitude in 1\arcs aperture is consistent with assigning
[OII] to this line; it is not consistent with Ly $\alpha$.

\noindent
{\bf CENSORS 34}: The spectrum has a complete absorption band at about
9200\AA\ with width \squig~~100\AA. This is not identified with any known
spectral feature that the authors are aware of. A redshift is assigned on
the basis of its $K$--band magnitude.

\noindent
{\bf CENSORS 36}: A single emission line is observed. This is assumed to
be [OII] as this fits very well with the $K$--$z$ relation estimate for
the redshift.

\noindent
{\bf CENSORS 37:} Whilst this source has broad lines it is also at
relatively low redshift (0.5) and is optically faint, perhaps suggesting
that is better classified a BLRG. However it lies far off the $K$--$z$
relation and so is classified as a quasar.

\noindent
{\bf CENSORS 38:} This source shows clear broad emission lines. It was not
selected as a quasar on the basis of the optical image in Paper 1 because
the host object was not identified there (it was below the likelihood
threshold). However, it is an unresolved optical ID. It is therefore a
quasar.

\noindent 
{\bf CENSORS 42:} This single line is clear in the 2D image. With a
$K$--band magnitude limit of 18.8 this is more likely to be Ly $\alpha$ or
CIII] than [OII].  The $K$--band magnitude might suggest a mild preference
for assigning \lya~as the line ID, however if that were the case there are
several other lines that might be expected at higher wavelengths
(e.g. CIV, HeII or CIII]) On the other hand, if the line were CIII], then
the next bright line would be [NeV], which would coincide with a sky
feature at 7700\AA.  Mg II would also be missing, but this is not always
detected in every case (see CENSORS 20).  Taking the line as CIII]: $z =
1.254$.

\noindent
{\bf CENSORS 49:} As with CENSORS 37 this may be better classified a BLRG,
rather than a quasar, and as it lies consistently on the $K$--$z$
relation, of \cite{Willott03KZ}, this is the classification it is given.

\noindent 
{\bf CENSORS 50:} There is a possible single line at 5882\AA, however it
is not clear from the 2D image whether this is in fact an emission line or
if it is a residual associated with sky lines.  There is also a faint line
at 4826\,\AA.  This is mildly visible in the 2D spectrum.  To be roughly
consistent with the $K$--band magnitude, which estimates the redshift as
$z \simeq 1.6$, the 4826\,\AA~line (which is more likely of the two to be
a genuine detection) will be at rest wavelengths of $\approx 1850$\AA.
The strongest line in this range is CIII] at 1909 giving $z = 1.528$.
This matches the 5882\,\AA~up to be CII] and would put HeII at 4145\,\AA.
Whilst HeII is not clearly detected in the 2D image, there are hints of
signal in the spectrum which add weight to this spectroscopic redshift.

\noindent
{\bf CENSORS 52:} Similarly to CENSORS 49, the line widths have large
errors and the source is quite faint. As it lies in good agreement with
the $K$--$z$ relation of \cite{Willott03KZ} it is taken to be a broad line
radio galaxy rather than a quasar.

\noindent 
{\bf CENSORS 55:} This low S/N detection of the Ca H and K and the
$G$--band absorption features is consistent with this source's K band
magnitude of 17.19(09) (1\arcsec aperture).

\begin{figure}
\centerline{\psfig{file=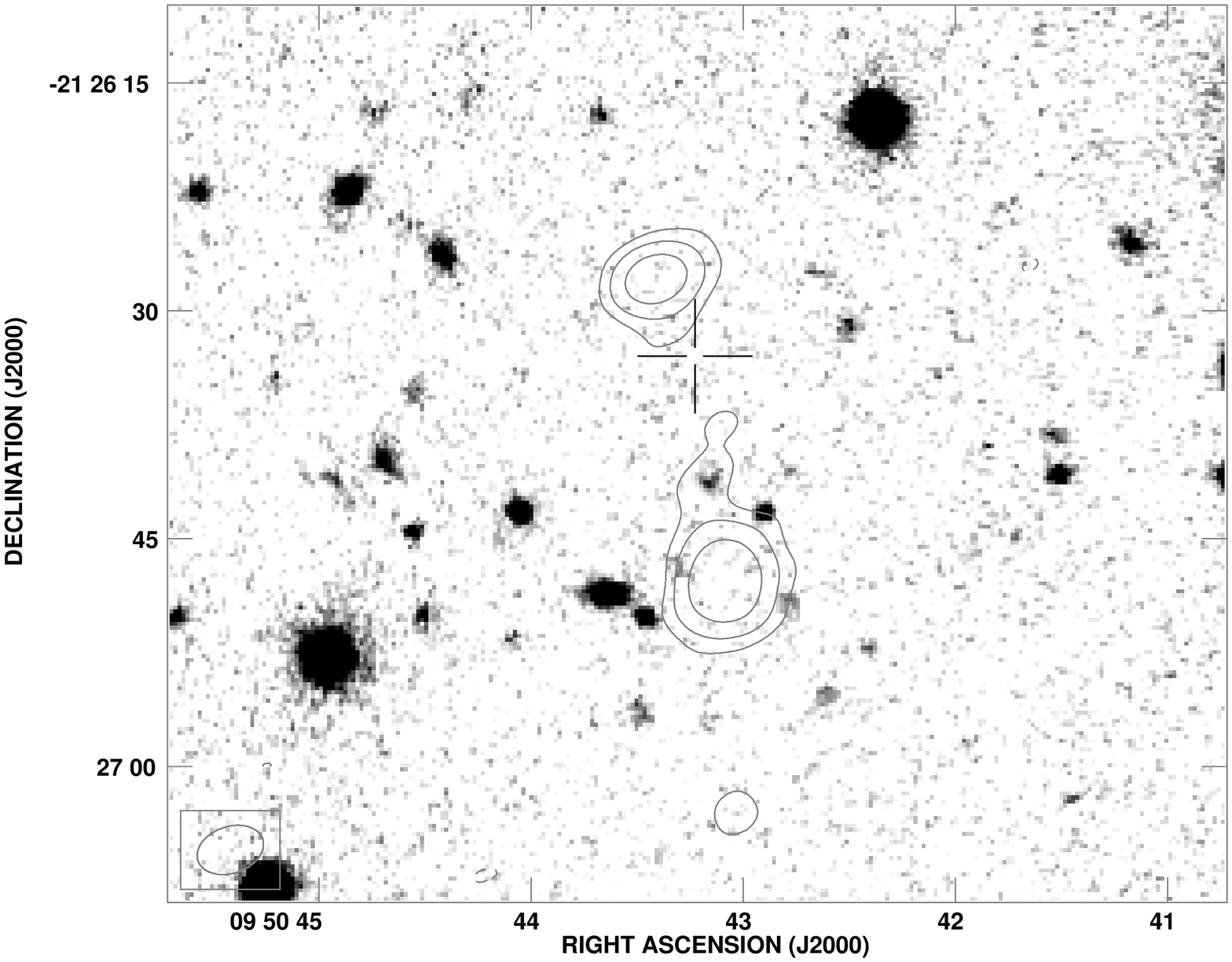,angle=0,width=6cm}}
\caption{The greyscale presents the $I$--band image of CENSORS 56. The
crosshairs indicate the position of the emission line detected in
spectroscopic observations of the source.\label{cen56linepos}}
\end{figure}

\noindent
{\bf CENSORS 56}: The target was a galaxy detected in both the $I$ and
$K$--bands between the two radio lobes of this source. The target spectrum
shows no features with which its redshift may be estimated. However the
slit was aligned with the northern radio lobe and a single emission line
was clearly detected there. The position of the line is shown in Figure
\ref{cen56linepos} and is 09 50 43.3 $-$21 26 32.4 (J2000) and it is
possible that this is in fact a very faint host galaxy associated with the
radio source. This line is assumed to be [OII]. At such a redshift it is
reasonable that no other equally strong would be seen in the spectrum. If
the line were [OIII] then [OII] might also be expected.

\noindent 
{\bf CENSORS 63:} There is a faint detection of a 4000\,\AA~break in this
spectrum from 2dF. This provides a redshift of 0.314.  This has been
confirmed in the blue spectrum from ISIS on the WHT which finds a break
corresponding to $z = 0.302$.  Whilst this is not a strong detection, this
redshift is consistent with the $K$--$z$ relation estimate of 0.3. The
latter redshift estimate is chosen as the 2df spectrum suffers from not
being flux calibrated. This object was classified as a quasar candidate on
the basis of both its stellaricity and colour in Paper 1.  However its
colour may not necessarily suggest that it a quasar (its $B$--band
magnitude is very close to the survey limit). In addition no emission
lines are seen in the spectrum. It is a radio galaxy.

\noindent 
{\bf CENSORS 65:} This low S/N detections of the Ca H and K and the G-band
absorption features are consistent with this source's $K$--band magnitude
of 17.77(16) (1\arcsec aperture).

\noindent
{\bf CENSORS 70}: In Papers 1 and 2 the host galaxy remained unidentified
from a choice of five candidates. Candidates B and E (as described in
Paper 2) were considered the most likely hosts on the basis of their
$I$--$K$ colours. A long slit was placed on both candidates B and E. The
resulting spectra showed that B was a late type star, but that E has both
the [OII] and [OIII] emission lines and thus harbours an active
nucleus. Candidate E, with a position of 09 48 10.6 -20 00 58.56 (J2000),
is concluded to be the host galaxy.

\noindent 
{\bf CENSORS 71:} This single line is assigned to Ly $\alpha$ because it
is very extended in the 2D image, as shown in Fig. \ref{lya_fig}.

\noindent 
{\bf CENSORS 72:} This single line is assigned to Ly $\alpha$ because it
is very extended in the 2D image, as shown in Fig. \ref{lya_fig}.

\noindent 
{\bf CENSORS 82:} The target, as identified in Paper 1 from the EIS I band
imaging, is a star. A redshift lower limit is estimated on the basis of
its limiting $K$--band magnitude.

\noindent 
{\bf CENSORS 88:} This single line is convincing in the 2D image despite
proximity to sky lines. Interestingly the target for these observations
was an $I$--band object which was not given as the host galaxy in Paper 1
as it was below the formal likelihood limit. This object was not observed
in the $K$--band. The $I$--band magnitude is 22.62 and the position is 09
45 20.95 $-$22 01 21.00. The line is taken to be [OII].

\noindent 
{\bf CENSORS 89:} There were two $K$--band identifications for this
source. The one at the centre of the extent of the radio source shows no
continuum/features; the eastern identification shows the spectrum
presented. The [OII] line and H and K absorption features are clear in the
2D image.

\noindent 
{\bf CENSORS 95:} As described in Section \ref{sbgals} this radio source
is due starformation rather than an active nucleus.

\noindent
{\bf CENSORS 100:} A spectrum was attempted for this source using the ESO
3.6m telescope, however no features were observed and it was decided this
was a more suitable target for the VLT. Despite the lack of features, it
can be concluded that this source is not a quasar since the necessary
bright emission lines would have revealed themselves in even a short
exposure.

\noindent
{\bf CENSORS 101}: There are absorption features which appear to match H
and K. However as this occurs at low S/N close to the beginning of sky
residuals, it would be comforting to have confirmation. The redshift
obtained from assuming these features are H and K is 1.043, which is
consistent with the $K$--$z$ relationship estimate of 0.985.

\noindent
{\bf CENSORS 102:} This galaxy is spectroscopically found at $z = 0.468$
and clearly has an old stellar population. This helps identify it as the
likely host despite the lack of emission lines.

\noindent 
{\bf CENSORS 105:} This source has very extended Ly $\alpha$ emission as
shown in Fig. \ref{lya_fig}.

\noindent
{\bf CENSORS 114:} Clear broad emission lines suggest this source as a
quasar. It was not classified as a quasar optically due to a $S/G$
slightly lower than the limit, but it has a blue colour so is likely to be
a quasar.

\noindent
{\bf CENSORS 116:} This source is interesting because of the strength of
the NeV line compared to the CIV line.  This rare enhancement of the
NV/CIV line ratio is discussed in \citeauthor{vanOjik94} 1994 for the case
of TX0211.  According to their work, the Ly $\alpha$/CIV line ratio of
CENSORS 116 is typical of high redshift radio galaxies, but, taking a
value of $> 1.6$, the NV/CIV is extremely enhanced. Following the
discussion of \citeauthor{vanOjik94}, this is likely due to enhanced
nitrogen abundance or shock processes.

\noindent {\bf CENSORS 118:} This single line has a very slight
confirmation suggesting it could be CIII] (1909) with CII] (2326) as the
extra line. However if this is the case then there is no detected [OII]
3727 emission which is unlikely, therefore it is assumed that this line is
Ly $\alpha$. This is also more consistent with its $K$--band (in a
1\arcsec aperture) of 19.82(15) than the CIII] assignment for that line
(or [OII] as a single line).

\noindent
{\bf CENSORS 119}: This radio source is elongated and the slit was aligned
along the radio axis. No redshift was measurable on the target spectrum,
however a single emission line was observed 1.4\arcs\ to the E of the
target. This is well within within the region of the radio source and it
is assumed that this emission line is representative of the host
redshift. Given the lack of any other emission lines, it is assumed that
this is [OII], resulting in a redshift of 1.484.

\noindent 
{\bf CENSORS 124:} This source has not been spectroscopically targetted,
but a redshift of 0.01559 has been found by \cite{cen124_z}. As described
in Section \ref{sbgals} this radio source is due to star formation rather
than an active nucleus.

\noindent 
{\bf CENSORS 126:} There were two likely host galaxies candidates in Paper
1. Of these the western source was targetted with 2dF.  This spectrum
revealed that this candidate was a star. The second candidate host galaxy
has not been spectroscopically targetted. It should be noted that the
radio morphology for this source is unclear and it is not clear how
realistic the remaining candidate is. However since the radio morphology
of this source is unclear, it can be said that it is unlikely to be a
quasar.

\noindent 
{\bf CENSORS 127:} This single line is assigned to [OII] for consistency
with its $K$--band magnitude of 17.46(21) in a 1.5\arcsec aperture.

\noindent
{\bf CENSORS 129:} Ignoring the \lya~line, only the CIV permitted line is
broad and this is a marginal case. Since the $K$--band magnitude is
consistent with the $K$--$z$ relation of \cite{Willott03KZ} is it assumed
to be a radio galaxy.

\noindent 
{\bf CENSORS 131:} There are no confirming lines or features for this
4000\AA~break. The $K$--band magnitude is 16.00(21) in a 0.6\arcsec
aperture and so a redshift of 0.470 fits with the $K$--$z$ relation
expectation.

\noindent
{\bf CENSORS 133}: A single line is detected. Assuming it to be [OII]
gives a redshift of 1.335 which is consistent with the $K$--$z$
relationship estimate of 1.834. If the line were [OIII] the corresponding
redshift would be 1.737, however we would expect to see [OII] at the low
wavelength end of the spectrum. The redshift is taken to be 1.335

\noindent
{\bf CENSORS 135}: A single broad line is detected along with
continuum. Given the shape of the line it is assumed to be MgII, resulting
in a redshift of 1.316. This is consistent with the $K$--$z$ prediction
however it should be noted that the continuum is blue and so this
prediction is not appropriate. Since the continuum is slightly blue this
source is assumed to be a quasar in order to err on the side of caution.

\noindent 
{\bf CENSORS 136:} In Paper 2 the host galaxy for this source was
identified with a faint $K$--band galaxy which was offset in position from
the main radio lobe (to the SE). This was justified on the basis that
there were indications of a mild extension in the radio structure, seen as
two faint peaks south of the main lobe. The host galaxy was targetted with
FORS1 with LSS with a position angle for which the slit also passed
through the centre of the main radio lobe.  When the spectrum was
analysed, the previously identified host showed no signs of activity,
however two emission lines were detected at the position of the main radio
lobe. These lines are used to measure a redshift for the source of 0.629.

\noindent 
{\bf CENSORS 141:} This single line is assigned to Ly $\alpha$ in
agreement with the $K$--$z$ relation prediction for this source ($K$ =
19.57(20) in a 1\arcsec aperture).

\noindent 
{\bf CENSORS 146:} This source has not been spectroscopically targetted,
but a redshift of 0.02935 has been found by \cite{cen146_z}. As described
in Section \ref{sbgals} this radio source is due starformation rather than
an active nucleus.

\noindent 
{\bf CENSORS 148:} The targetted object was a star and so a redshift has
been estimated on the basis of its $K$--band magnitude.

\section{Spectra of CENSORS sources}

The associated gif files (spec1.gif to spec13.gif) show spectra of the
CENSORS sources for which a spectroscopic redshift has been successfully
obtained.

\end{document}